\documentclass[aps,notitlepage,amsmath,amssymb,superscriptaddress,longbibliography,twocolumn,pra,floatfix]{revtex4-2}

\usepackage{xcolor}
\usepackage{graphicx}

\usepackage{mathrsfs}
\newcommand{\ud}{\mathrm{d}}

\newcommand{\pder}[2]{\frac{\partial#1}{\partial#2}}
\let\Eps\varepsilon
\newcommand{\order}[1]{\mathcal{O}(#1)}

\DeclareMathOperator{\sn}{sn}

\usepackage[breaklinks=true]{hyperref}
\graphicspath{ {./figures/} {./}}

\begin{document}
\title{Recurrent nonlinear modulational instability via down-conversion in quadratic media
}
\author{Andrea Armaroli}

\author{Simone Ferraresi}

\author{Gaetano Bellanca}

\author{Stefania Malaguti}
\affiliation{Dept. Engineering, Università di Ferrara, via Saragat 1, 44122, Ferrara, Italy}
\author{Fabio Baronio}
\affiliation{Department of Information Engineering, University of Brescia and INO CNR, Via Branze 38, 25123 Brescia, Italy}
\author{Stefano Trillo}
\affiliation{Dept. Engineering, Università di Ferrara, via Saragat 1, 44122, Ferrara, Italy}
\email{stefano.trillo@unife.it}

\begin{abstract}
We investigate the induced modulation instability in second-harmonic generation beyond the early stage of the linearized growth of the modulation. We find a regime of recurrence (quasi-periodic conversion and back-conversion between the pump and the modulation) which is genuine of the parametric conversion process in quadratic media.
Such recurrence is mainly driven by a process of non-degenerate downconversion, showing no analogy to the cascading regime which mimics the cubic (Kerr) nonlinearities.
We consider two different steady states, i.e., a pure second-harmonic and a mixed fundamental/second-harmonic state. Both exhibit this dynamics, which we show to be amenable to a description in terms of reduced frequency-truncated models. The comparison with full numerical simulations of the starting model prove the validity and robustness of the reduced models in characterizing in a simple and elegant way a wide range of modulationally-unstable steady states.
\end{abstract}

\date{\today}
\maketitle

\section{Introduction}
Modulational instability (MI) is a universal phenomenon that entails the exponential growth of sideband pairs at the expense of a uniform strong background or pump wave \cite{BT,Tai86}. Understanding how the MI evolves beyond the initial amplification stage is a topic that has attracted a considerable interest in cubic media \cite{VanSim01,Mussot18,Pierangeli18,Goossens19,Van21}. In particular, the MI seeded by a sideband pair and ruled by the nonlinear Schr\"odinger equation (NLSE) exhibits repeated cycles of conversion and back-conversion between the pump and a cascade of harmonic sideband pairs, a process that can be regarded as a form of Fermi-Pasta-Ulam-Tsingou (FPUT) recurrence \cite{VanSim01,FPUT,Armaroli24}.  The recurrence ruled by the NLSE is particularly interesting since it follows a complex phase-plane structure featuring two coexisting qualitatively different types of recurrences, as observed in recent experiments \cite{Mussot18,Pierangeli18}. However, MI recurrence is neither requiring the underlying model to be integrable \cite{Conforti16,Yao22}, nor is it a prerogative of media with cubic nonlinearity. In this paper, we investigate nonlinear MI occurring via second-harmonic generation (SHG) in quadratic media.

Although MI occurring via SHG was predicted three decades ago \cite{TF95a,TF95b,Buryak95,Kennedy96,Trillo97} (see also \cite{Buryak02} for a review) and observed shortly afterwards, mainly in spatial experiments \cite{Fuerts97,Fang00,Schiek01,Delque11,Jab21}, but occasionally also in time \cite{Salerno03} or spatio-temporal domain \cite{Salerno04}, it is only more recently that the strongly depleted regime of induced MI has been demonstrated to be experimentally accessible and potentially describable in terms of simple formulas \cite{Schiek19,Schiek21,Deng22,Trillo23}. However, the regime analysed so far concerns the so-called cascading regime, where pumping at $\omega_0$ under strong phase-mismatched SHG mimics Kerr-like dynamics \cite{cascading,Buryak02}. In such a regime, the dynamics is well known to be described by an effective cubic NLSE. Therefore, the machinery associated to exact or perturbative solutions of the NLSE \cite{Akh87,GS18,Conforti20,Baronio17} provides a useful approach to the characterisation of the quadratic dynamics.

The aim of this paper is to show that a recurrent stage of MI takes place in a genuinely quadratic regime, which relies on induced MI from pumping strictly at $2\omega_0$ or in a mixed
eigenmode of SHG with prevailing component at $2\omega_0$.  At variance with the cascading case analyzed before, the present mechanism is effective also close to phase-matching of SHG. This gives rise to much shorter recurrence distances compared with the cascading regime. Indeed the new regime allows us to avoid the shortcoming inherent to the cascading regime, i.e., the fact that the effective cubic nonlinear coefficient turns out to scale as $|\Delta k|^{-1}$ \cite{cascading}, where the absolute SHG mismatch $|\Delta k|$ must be intrinsically large for the cascading regime to be valid.

The paper is organized as follows: in Section \ref{s1}, we recall the model equations for SHG, its bifurcation structures of the eigenmodes, as well as the outcome of the MI stability analysis.
In Section \ref{s2} we analyze the recurrent MI arising from a pump at $2\omega_0$.
In Section \ref{s3} we show that the recurrent regime persists across the bifurcation point where mixed $\omega_0-2\omega_0$ pumping eigenmode appears.
Finally in Section \ref{s4}, we summarize our results. 
The interested reader will find the details of the mathematical derivations in the three Appendices.

\section{Model equations} \label{s1}
We start from an electric field $E=E_1(Z,T) \exp(i k_{\omega_0}Z -i \omega_0 T) + E_2(Z,T) \exp(i k_{2\omega_0}Z -i 2\omega_0 T)+\mathrm{c.c.}$ involving slowly-varying envelopes $E_1(Z,T)$ and $E_2(Z,T)$ centered around fundamental frequency (FF) $\omega_0$ and second harmonic (SH) $2\omega_0$. We consider the following dimensionless SHG model \cite{Buryak02} 
\begin{equation} \label{eq:SHGsystem}
\begin{split}
		& i\frac{\partial u_1}{\partial z} -\frac{\beta_1}{2}\frac{\partial^2 u_1}{\partial t^2} + u_2 u_1^*=0,\\
		& i\frac{\partial u_2}{\partial z}-\frac{\beta_2}{2}\frac{\partial^2 u_2}{\partial t^2} + \delta k u_2 + \frac{u_1^2}{2} =0,
 \end{split}	
\end{equation} 
where  $z = Z/Z_{nl}$ and $t=T/T_0$ are the propagation and temporal coordinates in units  of nonlinear length $Z_{nl}=(\chi \sqrt{I_t})^{-1}$ and  $T_0=\sqrt{k_1'' Z_{nl}}$, respectively. Here $k''_1$ is the group-velocity dispersions (GVD) at FF, and $\beta_{1,2}=k''_{1,2}/|k''_1|$ are normalized GVD coefficients, $I_t$ is the total injected intensity and $\chi=\omega_0[2/(c^3\epsilon_0 n^2_{\omega_0} n_{2 \omega_0})]^{1/2} d^{(2)}$ is the nonlinear coefficient, and {$\delta k =\Delta k Z_{nl} = (k_{2\omega_0}-2k_{\omega_0}) Z_{nl}$} is the normalized mismatch. The normalized envelopes $u_{1}(z,t)=\sqrt{2} E_{1}(Z,T)/\sqrt{I_t}$,  $u_{2}(z,t)=E_{2}(Z,T) e^{-i\delta k z}/\sqrt{I_t}$ are introduced to have Eqs.~ \eqref{eq:SHGsystem} in Hamiltonian form, which is convenient for our approach based on a Fourier mode truncation. In terms of these variables the conservation of total intensity and Hamiltonian $H$, respectively, read as
\begin{equation}
\begin{aligned}
    I=&\int_{-\infty}^{+\infty} \ud t \left(|u_2(z,t)|^2 + \frac{|u_1(z,t)|^2}{2}\right) ;\\
    H=&\int_{-\infty}^{+\infty} \ud t  \left(\sum_{m=1,2} \frac{\beta_m}{2} |\partial_t u_m|^2 + \delta k |u_2|^2\right. 
    \\& \left.  +  \frac{u_2 (u_1^*)^2+u_2^* u_1^2}{2}\right).
\end{aligned}
\end{equation}
Since Eqs.~\eqref{eq:SHGsystem} are not integrable, the nonlinear regime of MI must be described by approximate methods.

{Equations~\eqref{eq:SHGsystem} implicitly assume that higher-order mixing yielding frequencies such that $3\omega_0,4\omega_0,\ldots$ are highly mismatched and hence negligible. We further assume,} for the sake of simplicity, group-index matching, which can be achieved in bulk \cite{DiTrapani98} or through more modern dispersion engineering in lithium niobate on insulator platform \cite{Jankowski20,Kumar22}. We also point out that our results can be directly extended also to the diffractive case described by Eqs.~\eqref{eq:SHGsystem}, with $\beta_1=2\beta_2=-1$ \cite{Buryak02,Schiek19}.

\begin{figure}[t!]
\begin{center}
\includegraphics[width=7cm]{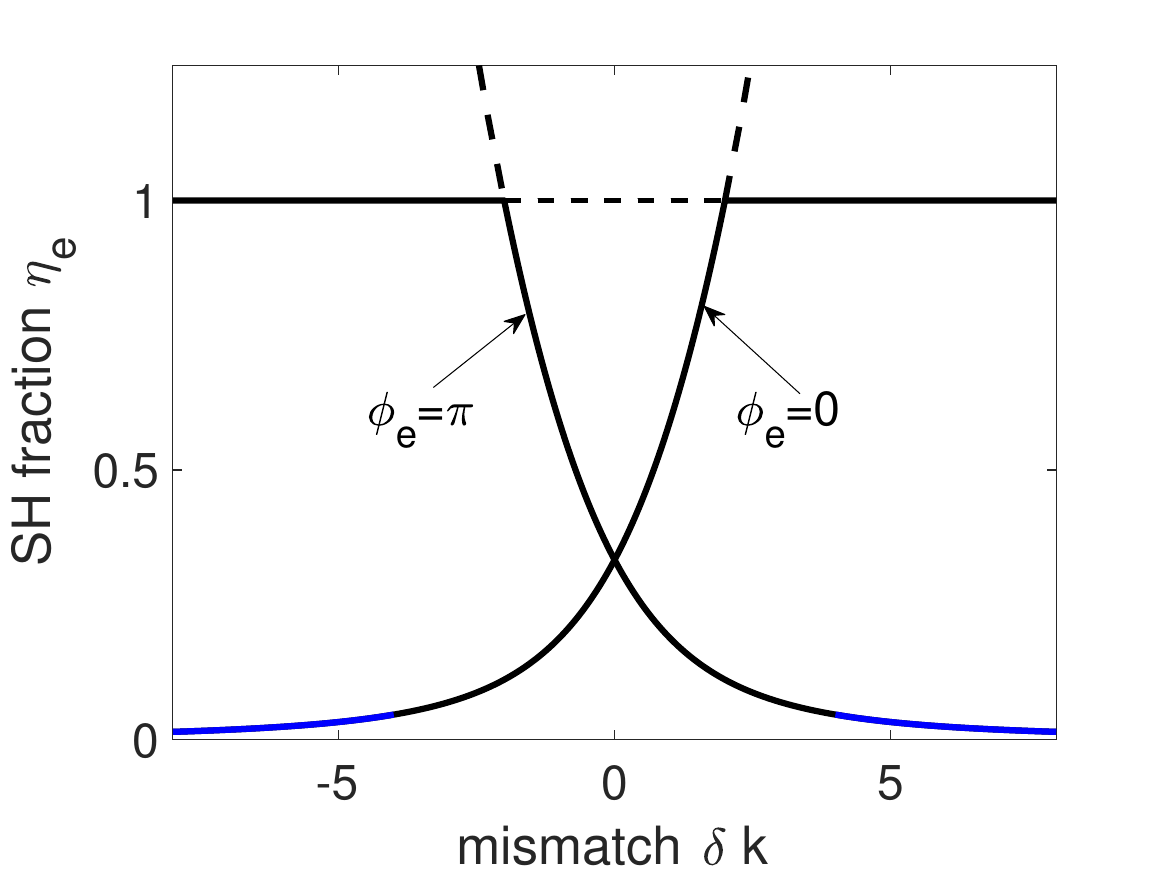}
\caption{Bifurcation structure of background (stationary) eigenmodes: SH intensity fraction $\eta_\mathrm{e}=|u_{20}|^2$ of equilibria vs.~SHG mismatch $\delta k$.
Solid and dashed lines indicate stable (center-type) and unstable (saddle-type) equilibria, respectively. The (saddle) branches with $\eta_\mathrm{e}>1$ are physically inaccessible. { Stable mixed FF-SH eigenmodes branches with different locked phases $\phi_e = 0$ and $\phi_e=\pi$ exist for $\delta k <2$ and $\delta k >-2$, respectively. The tails of these branches, i.e., for $|\delta k| \gtrsim 4$ and $\eta_\mathrm{e} <0.05$ (blue color traits), correspond to the cascading or Kerr-like regime responsible for a recurrent MI of different type compared with that discussed in this paper \cite{Trillo23}. Such Kerr-like MI is ruled by an effective focusing NLSE obtained for FF normal GVD ($\beta_1=1$) over the right tail ($\delta k \gtrsim 4$) or FF anomalous GVD ($\beta_1=-1$) over the left tail ($\delta k \lesssim -4$).}
} 
\label{fig1}
\end{center}
\end{figure}

First, we recall the main results about MI. The MI analysis is performed by inserting in Eqs. (\ref{eq:SHGsystem}) the following general Ansatz \cite{TF95a}
\begin{eqnarray} \label{eq:ansatzMI} 
 \begin{aligned}
    u_{1}(z,t) &=& \left[ u_{10} +  b_{1s}(z) e^{i \Omega t} + b_{1i}(z) e^{-i \Omega t} \right]e^{i \mu_1 z},\\
    u_{2}(z,t) &=&  \left[ u_{20} +  b_{2s}(z) e^{i \Omega t} + b_{2i}(z) e^{-i \Omega t}  \right]e^{i \mu_2 z}, 
  \end{aligned}
 \end{eqnarray}
where $u_{10}\exp(i \mu_1 z)$ and $u_{20}\exp(i \mu_2 z)$ are the components of the stationary ($z$- and $t$-independent) nonlinear eigenmodes of SHG \cite{Trillo92a}, and $b_{js}, b_{ji}$ stand for the amplitude of the sideband pairs (subscripts $s-i$ indicate signal-idler pairs) with {normalized detuning $\Omega$ (correponding to real-world detuning $\omega_\mathrm{MI}=\Omega/T_0$ from $\omega_0$ (FF, $j=1$) 
and $2\omega_0$ (SH, $j=2$), respectively}. Equations \eqref{eq:SHGsystem} are then linearized around the nonlinear eigenmode, obtaining a linear system $\dot{{\bf b}} = {\bf M} {\bf b}$ (henceforth dot stands for $d/dz$) in the column vector of perturbation amplitudes ${\bf b}=[b_{1s} ~b_{1i}^* ~b_{2s} ~b_{2i}^*]^T$. Here ${\bf M}$ is a $4 \times 4$ matrix, whose eigenvalues with positive real part yield the MI gain \cite{TF95a}.  The MI pump $(u_{10}, u_{20})$ are equilibria or nonlinear eigenmodes of Eqs.~\eqref{eq:SHGsystem} with $\partial^2/\partial t^2=0$, and, from the dynamical viewpoint, can be only of two types: centers (stable) and saddle points (unstable).  These SHG eigenmodes are summarized in the bifurcation diagram of Fig.~\ref{fig1} \cite{Trillo92a}, which shows the SH fraction $\eta_\mathrm{e}=|u_{20}|^2$ of the eigenmode  (the relative FF fraction is $|u_{10}|^2/2=1-\eta_\mathrm{e}$) against the SHG mismatch $\delta k$. 
\begin{figure}[b!]
\begin{center}
\includegraphics[width=8.5cm]{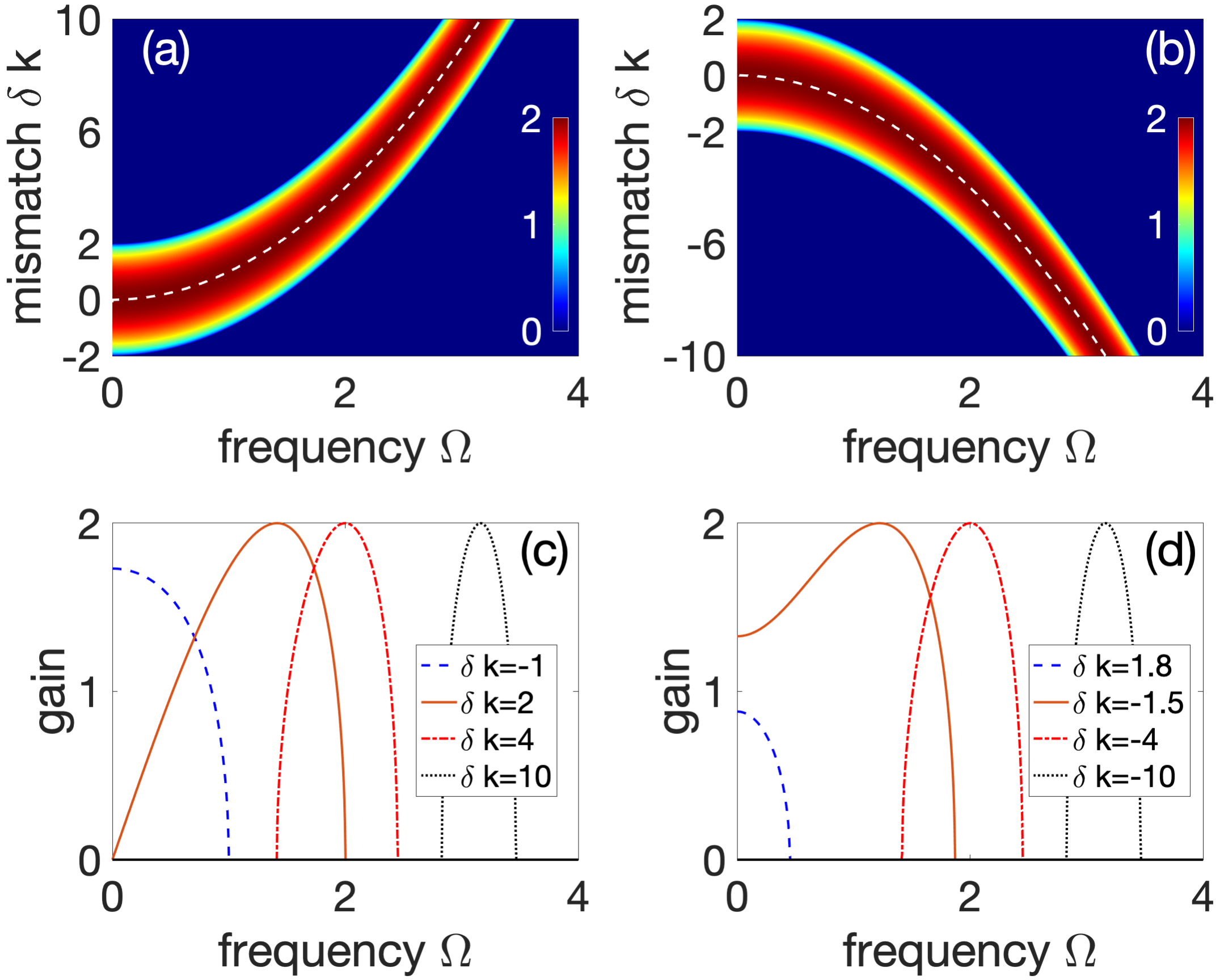}
\caption{(a,b) False-color plots of MI gain $g$ of the SH eigenmode ($u_{20}=1$) vs.~frequency $\Omega$ and SHG mismatch $\delta k$, in the (a) normal ($\beta_1=1$) and (b) anomalous  ($\beta_1=-1$) GVD regime. (c,d) Gain spectral profile $g(\Omega)$ sampled at different values of $\delta k$ in the (c) normal; (d) anomalous GVD regime. In (a,b) the dashed white curves gives the optimum (peak gain) frequency $\Omega_p=\sqrt{\delta k/\beta_1}$ corresponding to  phase-matching of 3WM downconversion.}
\label{fig2}
\end{center}
\end{figure}

As shown, the pure SH mode $\eta_\mathrm{e}=1$, corresponding to $u_{10}=0, u_{20}=1$ turns out to be a stable eigenmode for $|\delta k|>2$, whereas it becomes a saddle (unstable) for $|\delta k|<2$. At the two symmetric points $\delta k = \pm 2$, the SH mode bifurcates, exchanging its stability, with two branches of FF-SH mixed-type eigenmodes, which have their effective phase $\phi_e=(\mu_2-2\mu_1)z$ that remains locked either to $\phi_e=0$ or $\phi_e=\pi$. The latter are always stable (centers) in the range of interest $\eta_\mathrm{e} \le 1$ (only in the physically inaccessible regime corresponding to a SH fraction $\eta_\mathrm{e} >1$, they become saddle points).
When MI is pumped by the pure SH mode, the MI analysis, performed in the particular case $u_{10}=b_{2s,2i}=0$, in Eq.~\eqref{eq:ansatzMI}, yields the following MI gain
\begin{equation} \label{eq:gainSH}
g(\Omega)=2\sqrt{1 - \left( \frac{dk_3(\Omega)}{2} \right)^2};~~~\delta k_{3} (\Omega)\equiv \delta k - \beta_1 \Omega^2.
\end{equation}
Figure \ref{fig2}(a) shows a false-color plot of the gain in the plane $(\Omega, \delta k)$ in the case of normal GVD $\beta_1=1$, whereas gain curves sampled at different values of $\delta k$ are displayed in Fig.~\ref{fig2}(c). Clearly, Eq.~\eqref{eq:gainSH} shows that one gets identical gain curves with the transformation $\beta_1, \delta k \rightarrow -\beta_1, -\delta k$.
Indeed, in the anomalous GVD regime ($\beta_1=-1$), one simply gets the identical gain with opposite $\delta k$, as shown in Fig.~\ref{fig2}(b,d). In the following, we discuss in detail only the normal GVD case, keeping in mind that the results can be easily extended to the anomalous GVD regime by exploiting this simple symmetry.
In the normal GVD regime, MI takes place only above the bifurcation point $\delta k=-2$ [see Fig.~\ref{fig2}(a)]. In particular, in the range $-2<\delta k<0$, the gain peaks at $\Omega=0$ (see curve $\delta k=-1$ in Fig.~\ref{fig2}(c)), reflecting the nature of saddle point of the SH mode. In this case the width of the gain curve just measures the bandwidth of the parametric amplification of low frequency perturbations.
Conversely, for positive mismatch $\delta k>0$, a gain maximum emerges at finite frequency $\Omega_p=\sqrt{\delta k/|\beta_1|}$, that henceforth we call optimum frequency. This frequency corresponds to $\delta k_{3}(\Omega_p)=0$, i.e., to phase matching of the non-degenerate downconversion or three-wave mixing (3WM) photon interaction underlying the MI of the SH mode, {
namely $2\omega_0 \rightarrow (\omega_0+\omega_\mathrm{MI}) + (\omega_0-\omega_\mathrm{MI})$, recalling that $\omega_\mathrm{MI} \equiv \Omega T_0^{-1}$ is the dimensional detuning. Indeed $\delta k_3$ in Eq.~\eqref{eq:gainSH} is nothing but the normalized mismatch $\Delta k_{3}Z_{nl} \equiv \left[k_{2\omega_0} - k_{\omega_0+\omega_\mathrm{MI}} - k_{\omega_0-\omega_\mathrm{MI}} \right] Z_{nl}$, when $k_{\omega_0 \pm \omega_\mathrm{MI}}$ are expanded at second order around central frequency $\omega_0$.
Clearly, it is only the GVD at FF that affects the mismatch of such interaction}. 

In summary, for any $\delta k>0$, MI gain peaks at the frequency that realizes the phase-matching of 3WM, i.e. $\delta k_3(\Omega_p)=0$. As $\delta k$ increases, the gain band progressively narrows around the optimum frequency $\Omega_p$.

We also recall that a general feature of MI is that the peak gain frequency shifts with the intensity.
Indeed, in this case, the real-world detuning $\omega_{\mathrm{MI}p}=\Omega_p T_0^{-1}$ that corresponds to the optimum frequency, scales proportionally to the fourth root of the input intensity.

\begin{figure}[b!]
\begin{center}
\includegraphics[width=8.5cm]{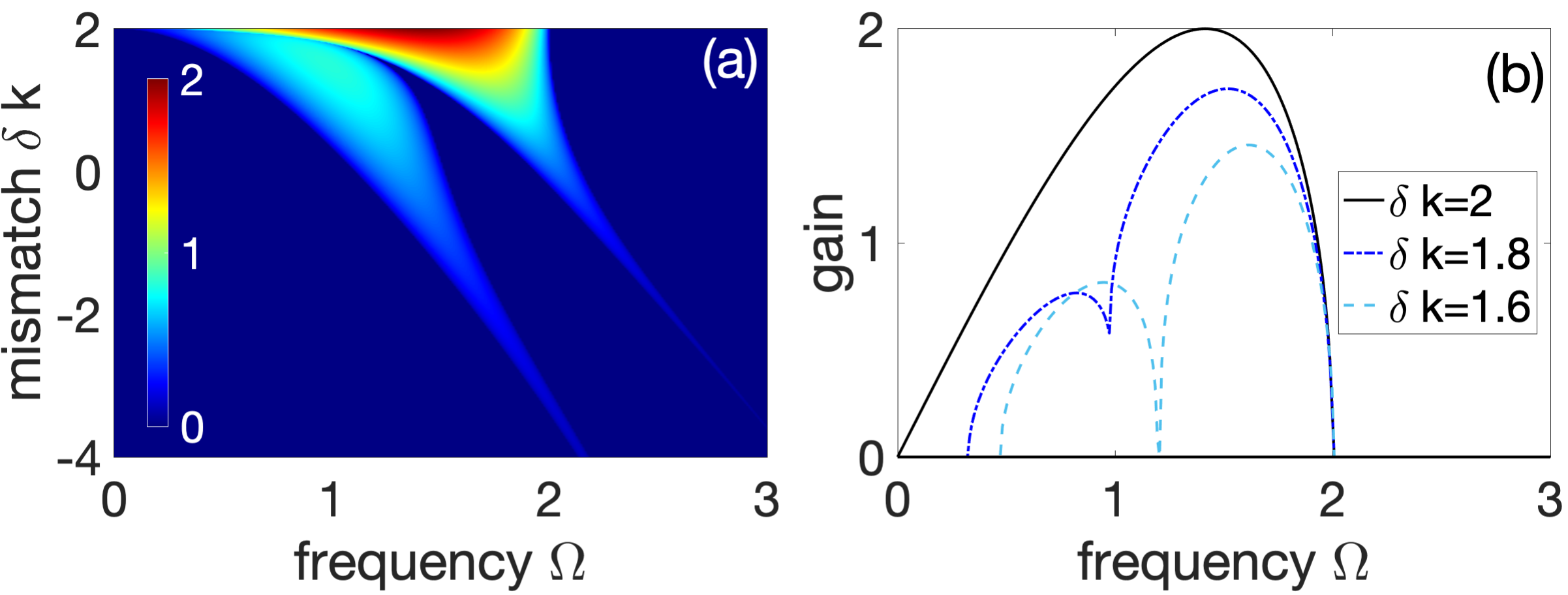 }
\caption{(a) MI gain $g$ of the mixed FF-SH phase-locked eigenmode with $\phi_e=0$ vs.~frequency $\Omega$ and SHG mismatch $\delta k$, in the normal GVD regime ($\beta_1=\beta_2=1$). (b) Gain spectral profile for $dk=1.6, 1.8, 2$ (bifurcation point in Fig. 1).
}
\label{fig3}
\end{center}
\end{figure}
By following instead the mixed-mode branch with $\phi_e=0$, the stability analysis gives, for the normal GVD case $\beta_1=1$, the MI gain displayed in Fig.~\ref{fig3}. As shown in Fig.~\ref{fig3}(a) there is a main branch at higher frequency that we consider henceforth since it gives considerably higher gain close to the bifurcation point $\delta k=2$, compared with the low-frequency branch (associated to complex conjugate eigenvalue pair). In particular, for the main branch, in the limit $\delta k=2$, the gain profile (see Fig.~\ref{fig3}(b)) is consistently identical to that of the SH mode in the same point. In this limit, the relative unstable eigenvector is composed only of symmetric sidebands around $\omega_0$ (i.e., $b_{1s}=b_{1i} \neq 0$ and $b_{2s}=b_{2i}=0$).
By decreasing the mismatch below $\delta k=2$, the gain rapidly drops, whereas the unstable eigenvector acquires non-vanishing components also for the sideband pairs around the SH. Furthermore, we point out that symmetry holds also for this mixed-mode case, so the same picture holds for the anomalous GVD with $\delta k \rightarrow -\delta k$, when considering the opposite branch with $\phi_e=\pi$.

In the following, our aim is to investigate whether the nonlinear regime of MI dominated by down-conversion, i.e., MI of the SH mode and mixed eigenmode close to the bifurcation point (i.e., the upper part the bifurcation diagram in Fig.~\ref{fig1} where $\eta_e \lesssim 1$), gives rise to recurrence. Indeed in this regime, we expect the dynamics to be dominated by few modes (the SH pump and a single sideband pair around FF, plus eventually the FF pump for the mixed mode eigenmode), for which a low-order truncation in Fourier space might be effective to describe quantitatively the dynamics. We remark that,  as far as the recurrent behavior of MI in SHG is concerned, a different regime exists that has been addressed in Refs.~\cite{Schiek19,Schiek21,Deng22,Trillo23}. This corresponds to the cascading regime, where the leading role is played by a FF stronger than the SH and one can successfully describe the dynamics in terms of an effective NLSE. This regime corresponds in the bifurcation picture of Fig.~\ref{fig1} to the tails (low $\eta_\mathrm{e}$) highlighted in blue. 
In particular, recurrences ruled by a focusing NLSE (fNLSE) occur either in the anomalous GVD regime, for negative $\delta k$ by exploiting the  $\phi_e=0$ eigenmode,
or for normal GVD and positive $\delta k$ on the $\phi_e=\pi$ branch.

\section{Recurrence pumped by the second-harmonic eigenmode}
\label{s2}

\begin{figure}[t!]
\begin{center}
\includegraphics[width=8.5cm]{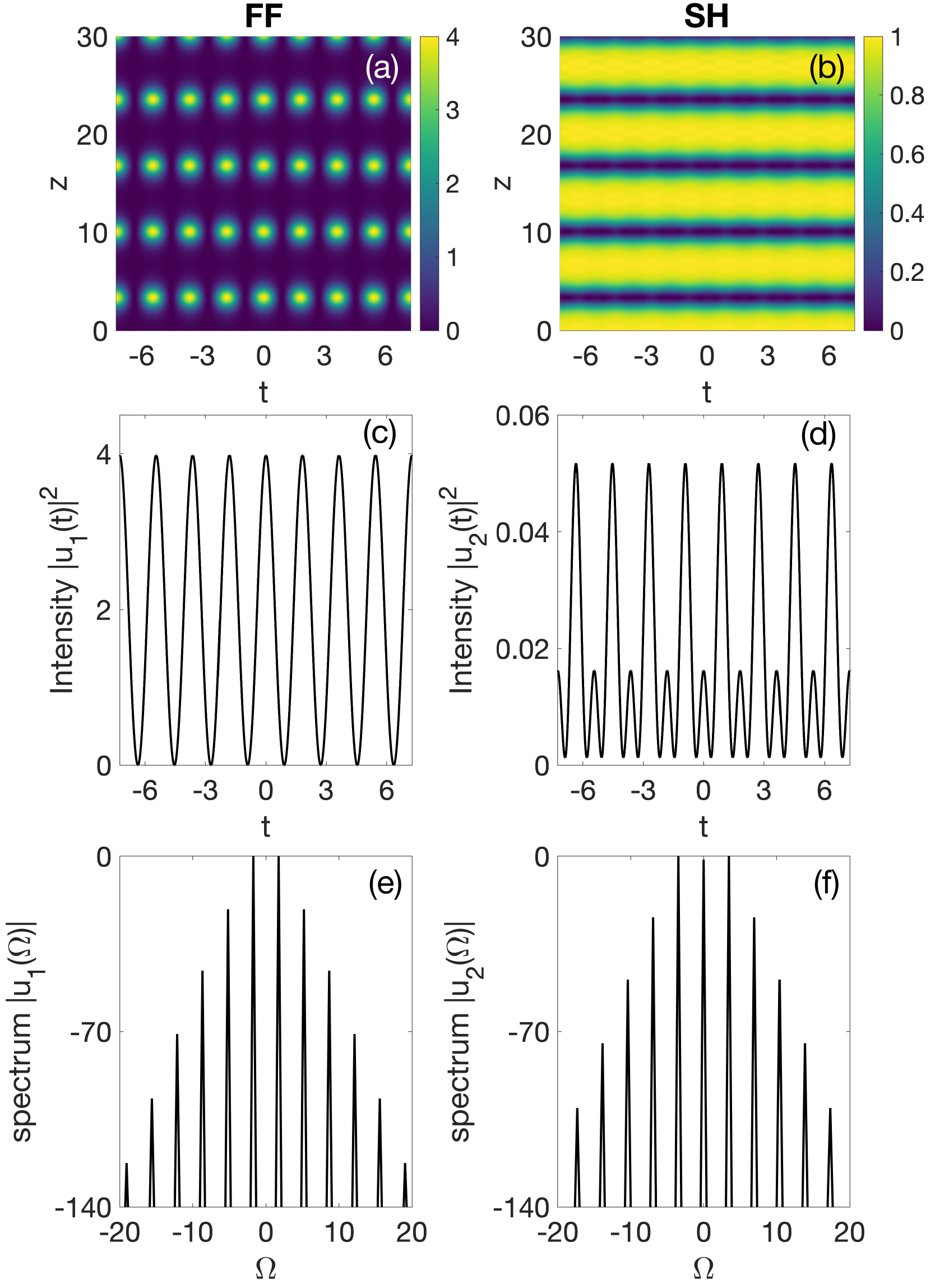}
\caption{
Evolution of FF (left) and SH (right): (a,b) false-color plots of the intensity dynamics in $(t,z)$ plane. (c,d) Intensity patterns vs.~time $t$ and (e,f) relative Fourier spectra vs. dimensionless detunings $\Omega=(\omega-\omega_0)T_0$ at FF and $\Omega=(\omega-2\omega_0)T_0$ at SH, sampled at first peak conversion distance $z=3.35$. Here, $a_1=0.1$, $\phi_1=0$, and $\beta_1=\beta_2=1$, $\delta k=3$, and the input modulation is at $\Omega=\Omega_p=\sqrt{3}$ such that $\delta k_3=0$.
}
\label{fig4}
 \end{center}
\end{figure}

{
In order to assess the long-term dynamics of the SH-pumped MI, we integrate numerically Eqs.~\eqref{eq:SHGsystem} with the following initial condition corresponding to seeded down-conversion from the SH mode
\begin{equation} \label{eq:inivalue}
    \begin{aligned}
    u_1(z=0)&=\sqrt{p}~a_1 \exp(i \phi_1) \left[\exp(i\Omega t) + \exp(-i\Omega t) \right]\\
    u_2(z=0)&=\sqrt{p},
\end{aligned}
\end{equation}
where $p \equiv 1/(1+a_1^2)$ 
in order to have unit normalized intensity. We find that the evolution follows, for sufficiently large mismatch ($\delta k >2$), recurrent, nearly periodic, evolutions as displayed in the example of Fig.~\ref{fig4}, where we take $\delta k = 3$, $\Omega=\Omega_p=\sqrt{3}$, $a_1 =0.1$ and $\phi_1 = 0$. In particular, a periodic trains of pulses (see Fig.~\ref{fig4}(c)) with zero mean in the complex amplitude is generated periodically in $z$, as also shown by the spectrum which exhibits two main lines at $\Omega=\pm \Omega_p$ (see Fig.~\ref{fig4}(e)). The underlying leading photon process is again $2\omega_0 \rightarrow (\omega_0+\omega_\mathrm{MI}) + (\omega_0-\omega_\mathrm{MI})$, which is exactly phase-matched for $\omega_\mathrm{MI}=\Omega_p T_0^{-1}$, i.e., $\Omega=\Omega_p$, or nearly phase-matched around that frequency. Sidebands with multiple harmonic frequencies are also generated through higher-order photon processes, leading to the nearly triangular spectrum of Fig.~\ref{fig4}(e). Clearly, at the apex of the pulse train generation, the SH shows its maximum depletion with weak residual modulation with main sidebands at $\pm 2\Omega$ ($2\omega_0 \pm 2\omega_\mathrm{MI}$ in real-world) arising from second-harmonic generation of primary sidebands at FF, as shown in Fig.~\ref{fig4}(f).}

\begin{figure}[t!]
\begin{center}
\includegraphics[width=8cm]{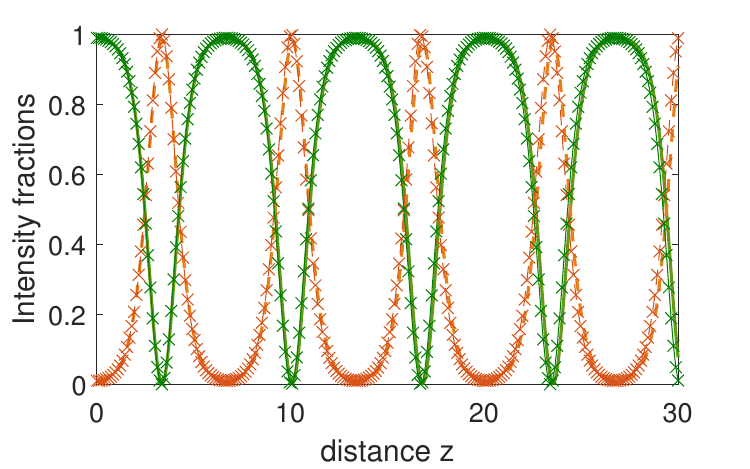}
\caption{Recurrent evolution {in space variable $z$} of intensity fractions of the SH (green) and sideband pair (orange), comparing those obtained from the PDEs Eqs.~\eqref{eq:SHGsystem} (thick solid and dashed lines) with those from the 3WM truncation [Eqs.~\eqref{eq:Hred}-\eqref{eq:period}] (thin solid and dashed lines with crosses). Parameters and input are as in Fig.~\ref{fig4}.
}
\label{fig5}
 \end{center}
\end{figure}

In order to describe the regime illustrated in Fig.~\ref{fig4}, we can derive a system of three ODEs by truncating to the three main modal amplitudes, retaining their $z$-dependence in order to describe full depletion. To this end, we insert in Eqs. (\ref{eq:SHGsystem}) the following Ansatz
\begin{equation} \label{eq:ansatz3}
    \begin{aligned}
    u_1(z,t)&=u_{1s}(z) e^{i\Omega t}+u_{1i}(z) e^{-i\Omega t}\\
    u_2(z,t)&=u_{20}(z),
\end{aligned}
\end{equation}
where $u_{20}$ and $u_{1s,1i}$ are the complex amplitudes of the SH modes and sidebands around FF, respectively. By neglecting all other generated frequencies, we obtain the following three-wave mixing (3WM) reduced description
\begin{equation} \label{eq:3wm}
	\begin{aligned}
		-i \dot{u}_{1s} & =\frac{\beta_1 \Omega^2 }{2}u_{1s}  + u_{20} u_{1i}^* 
        &=\frac{\partial H_3}{\partial u_{1s}^*},\\
		 -i\dot{u}_{1i} &= \frac{\beta_1 \Omega^2}{2} u_{1i} + u_{20} u_{1s}^* 
         &=\frac{\partial H_3}{\partial u_{1i}^*},\\
		 -i \dot{u}_{20} &= \delta \!k u_{20} +u_{1s} u_{1i}
         &=\frac{\partial H_3}{\partial u_{20}^*}.\\
	\end{aligned}	
\end{equation} 
where  $H_{3}=H_{3}(u_{20},u_{1s},u_{1i},u^*_{20},u^*_{1s},u^*_{1i})$ is the conserved finite-dimensional Hamiltonian which explicitly reads
\begin{equation} \label{eq:H3}
H_{3}= \frac{\beta_1 \Omega^2}{2} \sum_{j=s,i} |u_{1j}|^2  + \delta k  |u_{20}|^2 + \left(u_{20} u_{1i}^* u_{1s}^* + \mathrm{c.c.}\right),
\end{equation}
where c.c.~denotes the complex conjugate of the preceding terms.
As shown in Appendix A, this system is reducible to a single degree-of-freedom Hamiltonian system in terms of two conjugated variables $\psi(z)=\phi_{20}(z)-\phi_{1s}(z)-\phi_{1i}(z)$  (effective phase) and $\eta_{20}(z)\equiv |u_{20}|^2$ (SH fraction of total intensity).
For the sake of simplicity, henceforth we consider symmetric sidebands (i.e., $u_{1s}=u_{1i}$), which yield the following reduced system
\begin{equation} \label{eq:Hred}
 \begin{gathered}
    \dot\eta_{20} = -\pder{H_\mathrm{r}}{\psi};\; \dot\psi = \pder{H_\mathrm{r}}{\eta_{20}}\\
    H_{r}(\eta_{20},\psi)=\delta k_{3} \eta_{20} + 2\sqrt{\eta_{20}} (1- \eta_{20}) \cos\psi,
 \end{gathered}
\end{equation}
where the effective single parameter $\delta k_{3}=\delta k - \beta_1 \Omega^2$ is already defined in Eq.~\eqref{eq:gainSH}.

{
The seeded down-conversion corresponding to the launching condition in Eqs.~\eqref{eq:inivalue} is described by the solution of Eqs.~\eqref{eq:Hred} with initial value $\eta_{20}(z=0)=p$ and $\psi(z=0)=-2\phi_1$. Such solution can be written in terms of Jacobian elliptic sine ($\sn$) as \cite{elliptic}
}
\begin{equation} \label{eq:solution}
\eta_{20}(z)=\frac{a(b-c) \sn^2(\sqrt{a-c}~z |~k) - b(a-c)}{(b-c) \sn^2(\sqrt{a-c}~z |~k) - (a-c)},
\end{equation} 
{with $1-\eta_{20}(z)$ representing the complementary intensity fraction in the sideband pairs.
The spatial period of Eq.~\eqref{eq:solution} reads}
\begin{equation} 
z_p=\frac{2}{\sqrt{a-c}}K(k);\; \mathrm{with }\; k=\sqrt{\frac{b-c}{a-c}}.\label{eq:period}
\end{equation}
where K(k) is the complete elliptic integral of first kind, and $c \le b \le a$ are the ordered roots of a polynomial fixed by both the value of $\delta k_3$ and the initial condition (see Appendix A). 

\begin{figure}[ht!]
\begin{center}
\includegraphics[width=8cm]{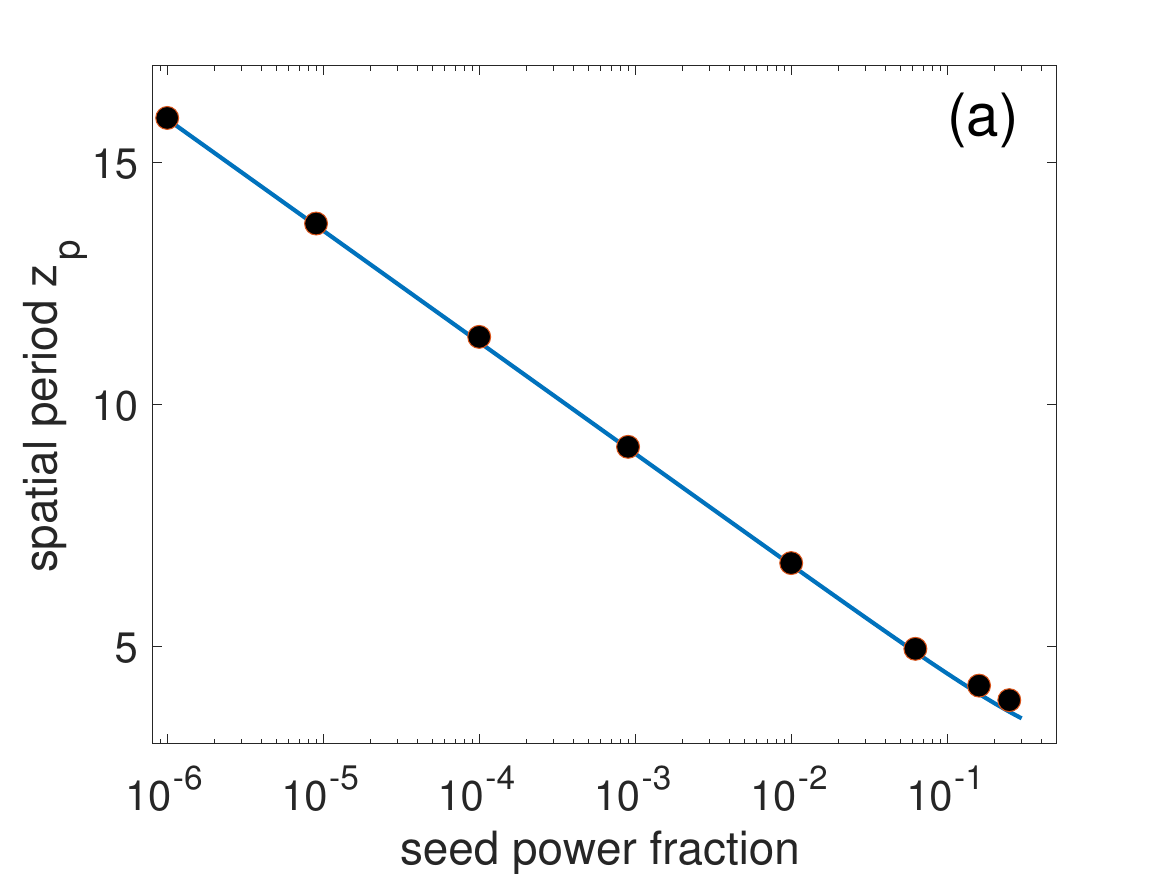 }
\includegraphics[width=8cm]{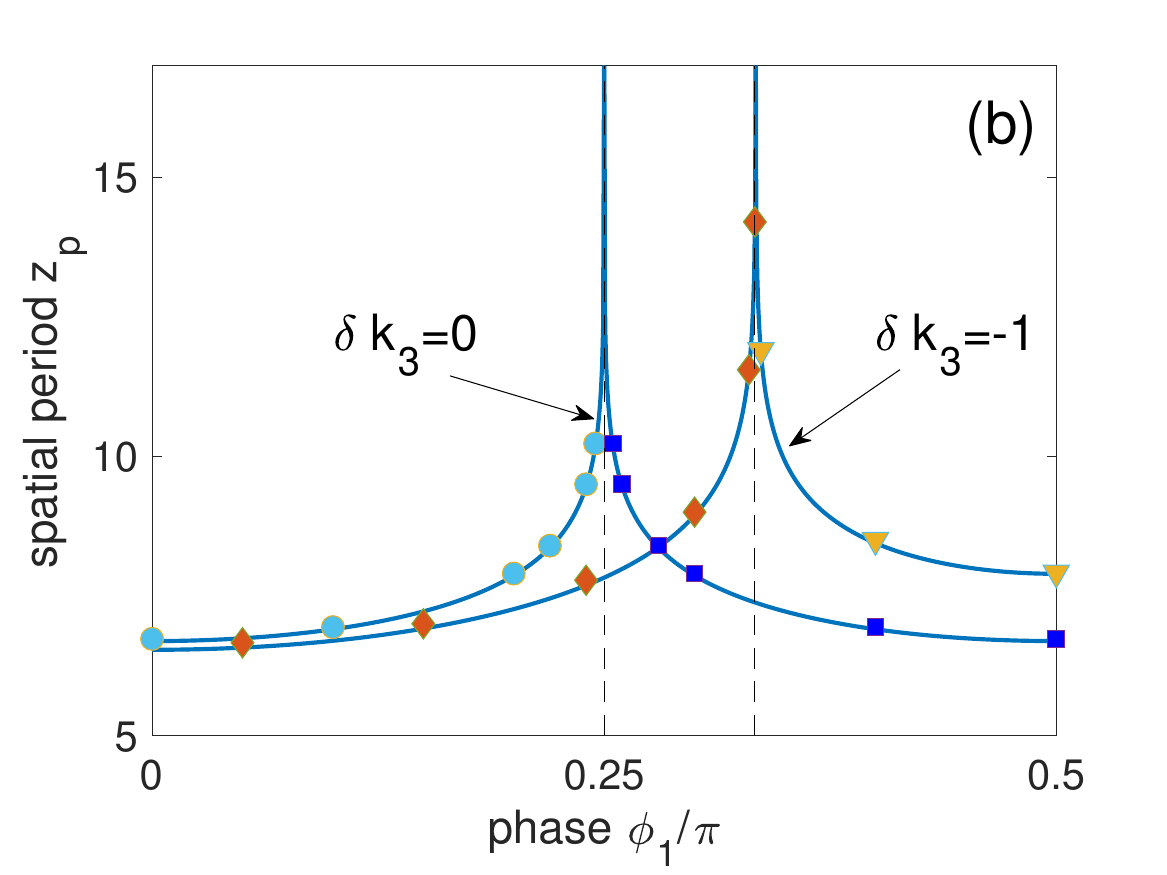 }
\caption{Period of MI recurrence comparing the results from full PDE simulations of Eqs.~\eqref{eq:SHGsystem} (dots) with the analytical estimate from Eq.~(\ref{eq:period}) obtained from 3WM truncation (solid line): (a) period vs.~power fraction of the seed for fixed phase $\phi_1=0$, $\delta k=3$, $\Omega_{p}=\sqrt{3}$; (b) period vs.~phase of seed $\phi_1$ for fixed $a_1=0.1$, and two values 3WM mismatch $\delta k_3=0$ ($\Omega=\sqrt{3}$) and $\delta k_3=-1$ ($\Omega=2$). Here $\beta_1=\beta_2=1$.}
\label{fig6}
 \end{center}
\end{figure}

The solution in Eqs.~\eqref{eq:solution}-\eqref{eq:period} gives an accurate description of the full PDEs dynamics as shown in the example of Fig.~\ref{fig5}, where we compare intensity fractions of the pump and the sideband pair from Eq.~\eqref{eq:solution} with the corresponding quantities extracted from the numerical integration of  Eqs.~\eqref{eq:SHGsystem} with initial value and parameters as in Fig.~\ref{fig4}.

{A systematic comparison of Eq.~\eqref{eq:period} with solutions of Eqs.~\eqref{eq:SHGsystem} is shown in Fig.~\ref{fig6}.} 

In particular, in Fig.~\ref{fig6}(a), we show the dependence of the period on the intensity of the sideband seed $a_1^2$ for fixed $\phi_1 =0$. It is apparent that Eq.~\eqref{eq:period}  provides an excellent estimate over several orders of magnitude of the sideband seed. The period shows also a marked sensitivity to the initial relative phase of the sidebands, as displayed in Fig.~\ref{fig6}(b), for fixed $a_1=0.1$.
In particular, the recurrence period $z_p$ predicted by the truncated model [Eq.~\eqref{eq:period}] is shown in Fig.~\ref{fig6}(b) as solid lines, for two different values of the 3WM mismatch, namely on-matching $\delta k_3=0$ (i.e., peak gain $\Omega=\Omega_p$) and off-matching $\delta k_3=-1$. Importantly, as shown, the period of the multiple conversion and back-conversion between the SH pump and the sidebands changes with $\phi_1$, and  tends to diverge as the sideband phase $\phi_1$ approaches (from either below or above) the critical value $\phi_{1s}=\frac{1}{2} \cos^{-1} (\delta k_3/2)$. Indeed, at this value of the phase, the evolution is expected to occur, in the limit of sufficiently small sideband amplitude, along a separatrix trajectory in the phase-plane associated to the Hamiltonian of Eq.~\eqref{eq:H3}. 
In particular, the separatrix connects the unstable manifold of the saddle equilibrium point $\eta_{20}=1, \psi=\psi_s=- 2\phi_{1s}$ to the stable manifold of the saddle with opposite phase ($\eta_{20}=1, -\psi_s=2\phi_{1s}$). Therefore, at this critical phase, the sideband pair are expected to asymptotically reconvert to the SH pump after a  single growth cycle {(in this sense this would be the genuine quadratic analog of the solution of the NLSE known as Akhmediev breather \cite{Akh87,Baronio17,Schiek19,Schiek21})}. The period extracted from numerical simulations of Eqs.~\eqref{eq:SHGsystem} and reported in Fig.~\ref{fig6}(b) by filled circles, turns out to be in good agreement with the analytical prediction from Eq.~(\ref{eq:period}), up to values close to the critical phase $\phi_{1s}$. However, the generation of multiple frequencies prevents the dynamics ruled by Eqs.~\eqref{eq:SHGsystem} to follow the separatrix of the 3WM truncation. Rather, we find that, as the critical phase $\phi_{1s}$ is approached, the rise of the period saturates and a multiple recurrent behavior sets in (evolutions not shown). 

\begin{figure}[ht!]
\begin{center}
\includegraphics[width=8.5cm]{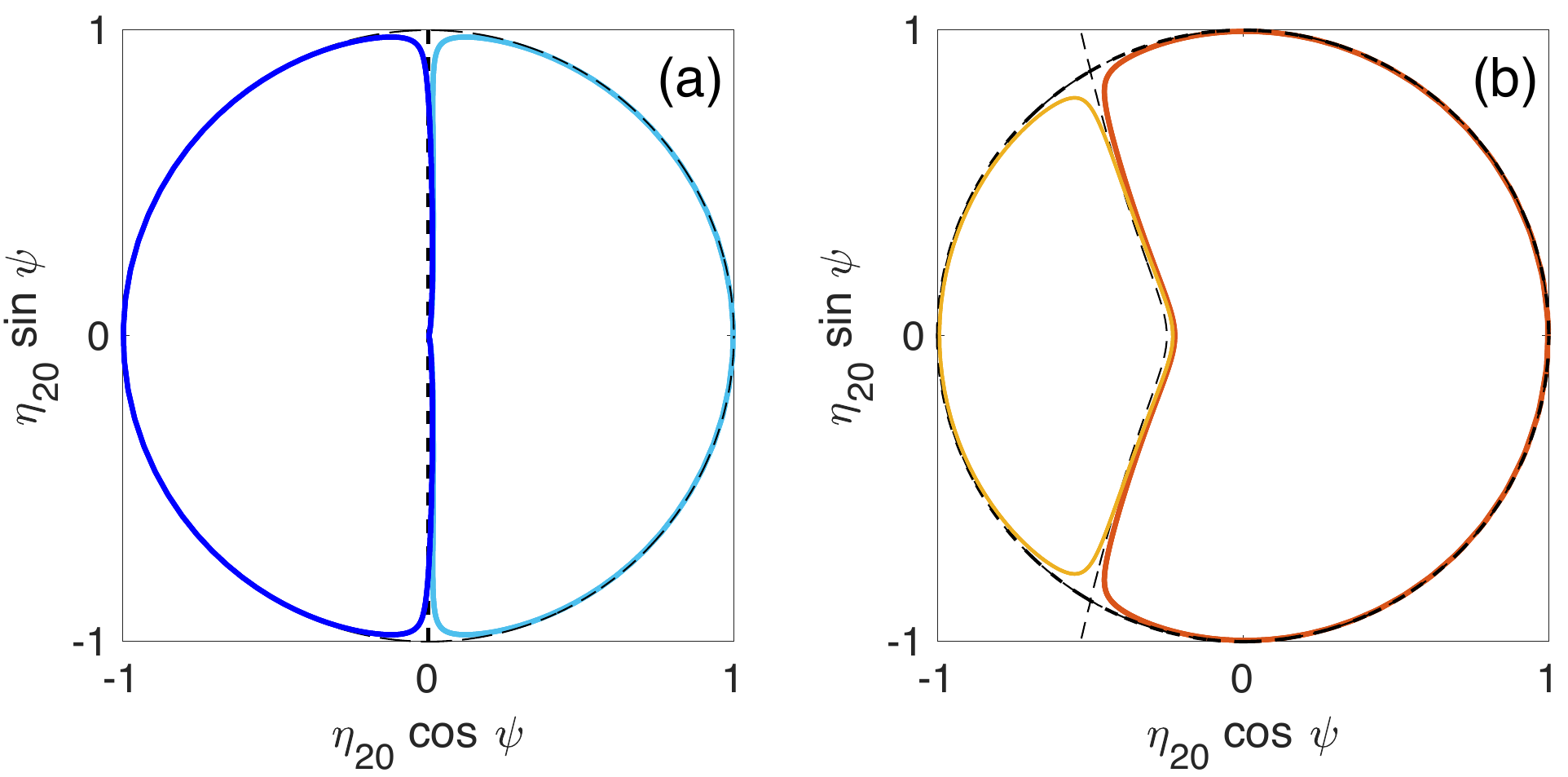}
\caption{Projections of PDE evolutions over the reduced phase-plane $(\eta_{20} \cos \psi, \eta_{20} \sin \psi)$ associated with the Hamiltonian reduction in Eqs. (\ref{eq:Hred}).
All trajectories are obtained with $\delta k=3$ and initial sideband amplitude  $c_1=0.1$,  (a)  Optimum frequency $\Omega=\Omega_p$ ($\delta k_3=0$), $\phi_1=0.22 \pi$ (cyan trajectory) and $\phi_1=0.28 \pi$ (blue trajectory); (b)  Non-optimum MI frequency $\Omega=2$ yielding 3WM mismatch $\delta k_3=-1$, $\phi_1=0.3 \pi$ (yellow trajectory) and $\phi_1=0.4 \pi$ (ochre trajectory).}
\label{fig7}
 \end{center}
\end{figure}

Nevertheless, the full PDE simulations exhibit a clear signature of the phase-plane structure associated with the Hamiltonian reduction in Eq.~\eqref{eq:H3}. In order to demonstrate this, in Fig.~\ref{fig7} we show a projection of the trajectories obtained from Eqs.~\eqref{eq:SHGsystem} onto the phase-plane $(\eta_{20} \cos \psi, \eta_{20} \sin \psi)$, as obtained for two values of $\phi_1$ across the critical phase $\phi_{1s}$ (the two trajectories are identified by the same color of filled circles used in Fig.~\ref{fig6}(b)). In particular, Fig.~\ref{fig7}(a), relative to the case $\delta k_3=0$, shows that the trajectories corresponding to $\phi_1>\phi_{1s}$ or $\phi_1<\phi_{1s}$, while presenting similar evolutions in terms of intensity fractions, evolve in the left or right half of the phase-plane, thus showing completely different phase dynamics. These two periodic orbits are on opposite sides of the separatrix which is the vertical bisector ($\psi=\pm \pi/2$). In the case $\delta k_3=-1$, shown in Fig.~\ref{fig7}(b), the two trajectories are still on opposite sides of the separatrix (dashed curve connecting $\psi=-\pi/3$ to $\psi=\pi/3$ on the unit circle $\eta_{20}=1$), hence exhibiting different phase evolutions. In this case it is important to notice that the shape of trajectories in the phase-plane is different on the two opposite sides of the separatrix, although the period of oscillation is almost the same. This is in stark contrast with the representation used for the NLSE, where trajectories inside and outside the separatrix are easily identified and exhibits period doubling \cite{Mussot18,Armaroli24}.

\begin{figure}[ht!]
\begin{center}
\includegraphics[width=8.5cm]{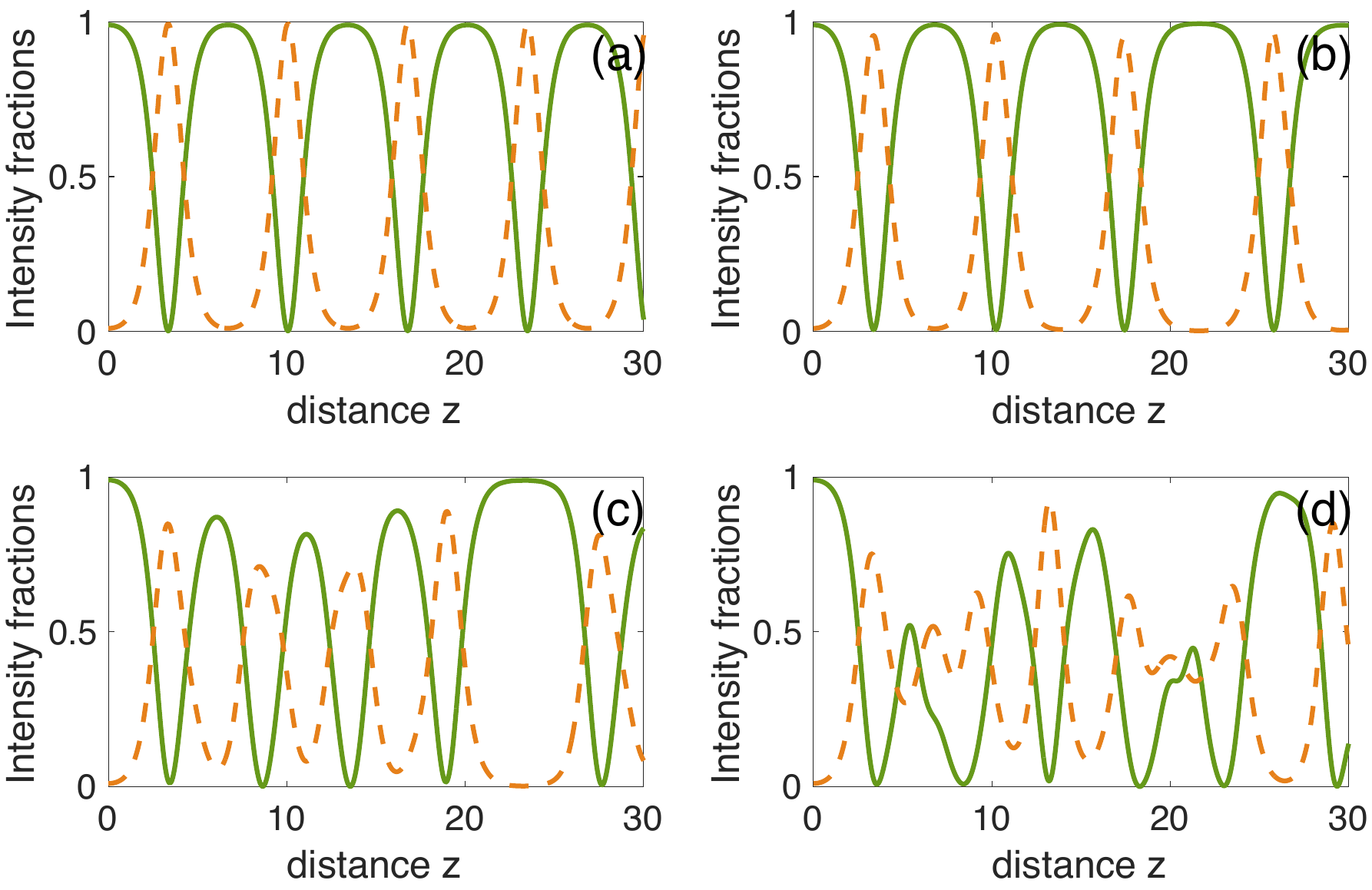}
\caption{Evolutions of intensity fraction of main frequency components SH (green) and sideband pair (orange) as obtained from the PDEs Eqs.~\eqref{eq:SHGsystem} for different SHG mismatch: (a) $\delta k=5$; (b) $\delta k=2$; (c) $\delta k=1$; (d) $\delta k=0.5$.
Here $\beta_1=\beta_2=1$, and he initial modulation has $a_1=0.1$
$\phi_1=0$, and $\Omega = \Omega_p = \sqrt{\delta k}$.}
\label{fig8}
 \end{center}
\end{figure}

A natural question is whether the 3WM truncation gives a good description of the recurrent dynamics in the whole range of positive $\delta k$ where MI exhibits peak gain at the finite frequency $\Omega_p$. It turns out that the agreement is excellent at large mismatches but still remains quite good up to the bifurcation point $\delta k=2$. When $\delta k$ decreases below this point, the PDEs exhibit progressively less regular recurrence cycles until, eventually, a rather irregular spatial behavior emerges. This is apparent in Fig.~\ref{fig8} where we show the spatial evolutions of the modes obtained by numerical integrations of the PDEs for four different values of mismatch $\delta k=10, 2, 1, 0.5$. The reason why the dynamics looses its regularity is the fact that the spontaneous amplification of low frequency noise components sets in due to the non-vanishing MI gain at $\Omega=0$ and nearby frequencies. The gain at low frequencies indeed grows larger as the mismatch decreases below the value $\delta k=2$ (see Fig.~\ref{fig2}(a)) and the gain curve flattens as $\delta k$ approaches zero. This causes the regularity of the dynamics to be spoiled due to the competing growth of low frequency noise components.
 
We point out, however, that, below $\delta k=2$, a new mixed-mode stationary branch appears, which continues to exhibit (sufficiently close to the bifurcation point) a rather regular recurrence, as illustrated in the following section.   

\section{Recurrent down-converted MI from phase-locked eigenmode}
\label{s3}

At $\delta k = 2$, a phase-locked mixed FF-SH eigenmode bifurcates from the SH one, and these two eigenmodes coexist for $\delta k \le 2$. In other words, MI can also develop on top of a two-color pump that possesses both components  $u_{10},u_{20}$.
In this case, we integrate Eqs.~\eqref{eq:SHGsystem} with seeded initial conditions 
\begin{equation}
    \begin{aligned}
    u_1(z=0)&=\sqrt{p}\left\{ \sqrt{2(1-\eta_\mathrm{e})}
    \right.
    \\
    &\left. + a_1 \exp(i \phi_1) \left[\exp(i\Omega t) + \exp(-i\Omega t) \right] \right \}\\
    u_2(z=0)&=\sqrt{p}\sqrt{\eta_\mathrm{e}},
    \end{aligned}
\end{equation} with $p \equiv 1/(1+a_1^2)$, as in Eq.~\eqref{eq:inivalue} and $\eta_\mathrm{e}$ the SH intensity fraction of the pump eigenmode of SHG obtained in \cite{Trillo92a}.

\begin{figure}[htpb]
\begin{center}
\includegraphics[width=8.5cm]{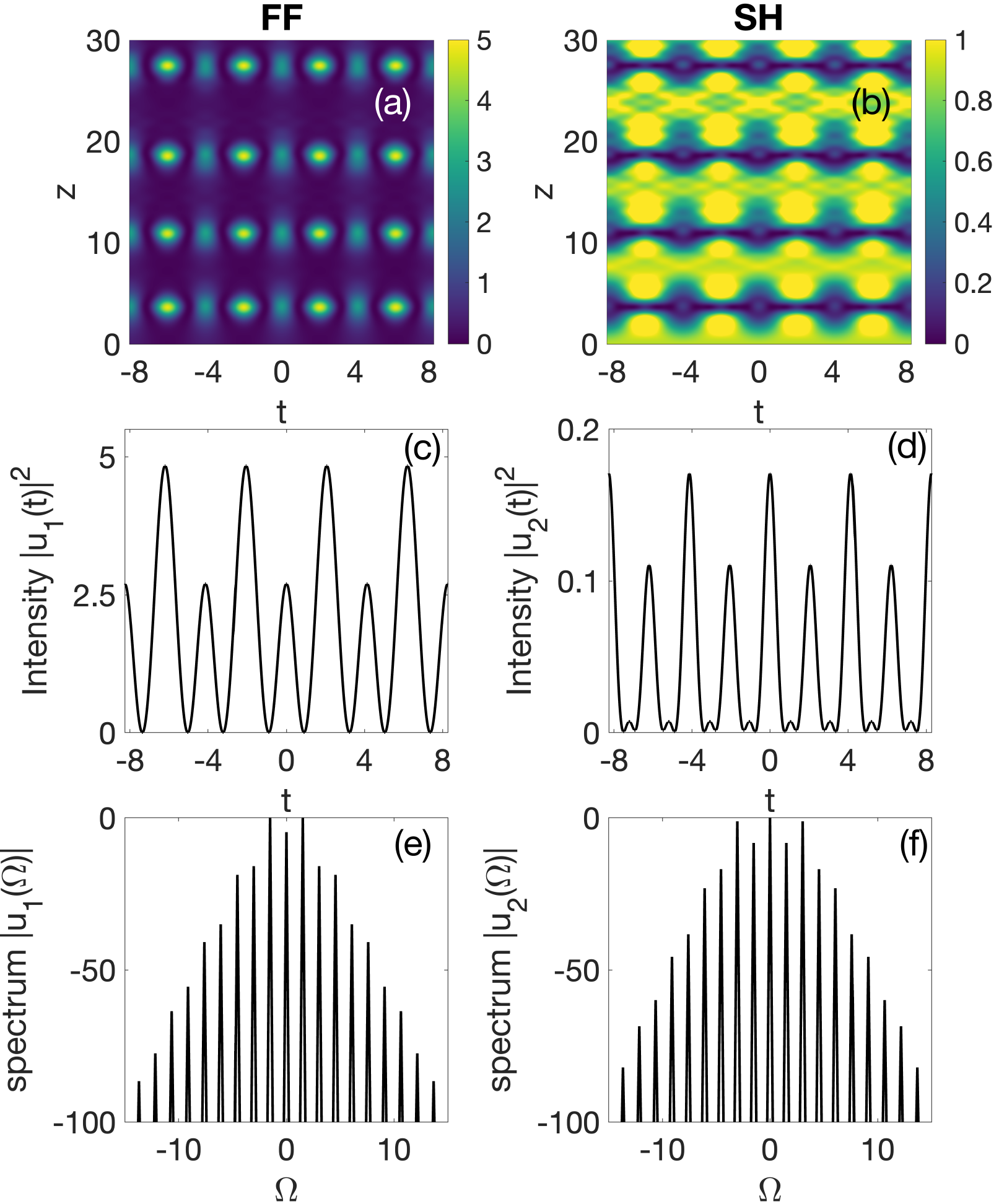}
\caption{As in Fig.~\ref{fig4} when the initial condition is the perturbed in-phase ($\phi_e=0$) mixed-mode at $\delta k=1.8$. Here the modulation frequency is $\Omega=\Omega_p=1.52$. Panels (c--f) are extracted at $z= 3.65$. The initial perturbation is the same of Fig.~\ref{fig4}. 
}
\label{fig9}
 \end{center}
\end{figure}

\begin{figure}[htpb]
\begin{center}
\includegraphics[width=9cm]{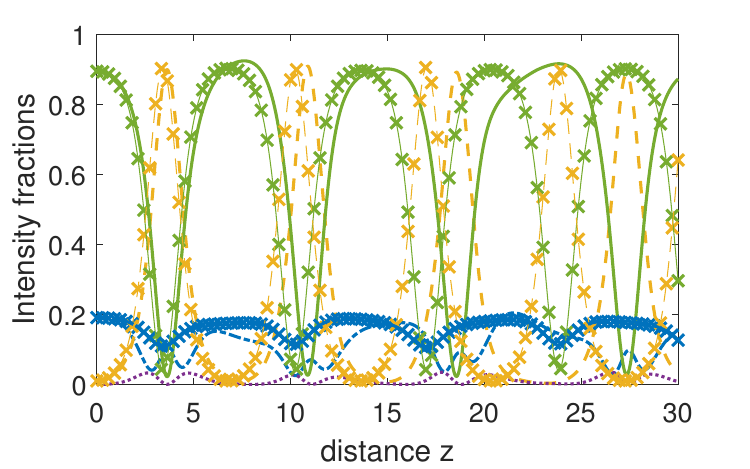}
\caption{Evolutions of intensity fraction of modes $|u_{20}|^2$ (solid green lines), $|u_{10}|^2$ (dash-dotted blue), $|u_{1s}|^2=|u_{1i}|^2$ (dashed orange lines). We compare the results of Eqs. (1) (thicker lines) and Eqs.~\eqref{eq:4wm} (thinner lines with crosses). {The dotted purple line stand for the SH sidebands $|u_{2s}|^2=|u_{2i}|^2$ neglected in Eqs.~\eqref{eq:4wm}.} Here $\delta k=1.8$, $\Omega=1.52$. The initial perturbation is the same of Fig.~\ref{fig5}. 
}
\label{fig10}
 \end{center}
\end{figure}

In Fig.~\ref{fig9} we show an example of recurring behavior for $\delta k=1.8$ at $\Omega=\Omega_p=1.52$, taking $a_1 = 0.1$ and $\phi_1=0$ as initial conditions. We notice in Fig.~\ref{fig9}(a,b) that the behavior appears to be recurring (obviously, both the FF and SH) with a period that slightly changes from one recurrence to the next. The intensities exhibit a rich dynamics, due to the complicated cascaded processes triggered in the nonlinear stage of MI. Since not only the SH but also the FF has an average component, the intensity of both of them oscillates with a period $\pi/\Omega_p$, similar to what was observed in Fig.~\ref{fig4}(d). This is apparent at the first focusing peak of the breathing dynamics ($z=3.63$), see Fig.~\ref{fig9}(c-d). Very interesting are also the spectral intensities, shown in Fig.~\ref{fig9}(e-f). {It is apparent that the components at $\pm \Omega_p$ around the FF are still the dominant ones in the MI process as predicted by the linear stability analysis recalled in Sec.~\ref{s1};  the spectral component at $\Omega=0$ corresponds naturally to the FF component of the mode, $u_{10}$. 
Conversely, sidebands at $\pm\Omega_p$ around SH (which now arise from non-degenerate up-conversion $\omega_0 + (\omega_0 \pm \omega_\mathrm{MI}) \rightarrow 2\omega_0 \pm \omega_\mathrm{MI}$ that involves the FF pump component) never grow significantly, as apparent in Fig.~\ref{fig9}(f). Instead, the second harmonic components of the primary sidebands at $\pm \Omega_p$ around FF, yielding $\pm 2\Omega_p$ around SH ($2\omega_0 \pm 2\omega_\mathrm{MI}$ in real-world units) are just few dB below the average component at SH. This is consistent with Fig.~\ref{fig9}(d), where oscillations reach almost the same level every $\pi/\Omega_p$. In this case, the spectra at both FF and SH remarkably depart from the triangular shapes characteristic of Fig.~\ref{fig5} or the nonlinear stage of MI ruled by the NLSE \cite{Mussot18}.}

In order to model this rich dynamics, which involves also a pump term at FF, we resort to a four-wave mixing (4WM) truncation that provides a reasonable description of the recurrent dynamics. To this end, we insert in Eqs.~\eqref{eq:SHGsystem} the following generalization of Eqs.~\eqref{eq:ansatz3} which accounts for the FF pump component
\begin{equation} \label{eq:ansatz4}
    \begin{aligned}
    u_1(z,t)&=u_{10}(z) + u_{1s}(z) e^{i\Omega t}+u_{1i}(z) e^{-i\Omega t},\\
    u_2(z,t)&=u_{20}(z),
\end{aligned}
\end{equation}
where we choose to neglect sidebands around SH, i.e., $u_{2s}=u_{2i}=0$, on the basis that the linearized analysis reveals that $u_{1s}$ and $u_{1i}$ are the dominant components of the unstable eigenvector. We obtain
\begin{equation}
	\begin{aligned}
		 -i\dot{u}_{10} &= u_{20} u_{10}^* 
         &=\frac{\partial H_4}{\partial u_{10}^*}\\
		-i\dot{u}_{1s} & =\frac{\beta_1 \Omega^2 }{2}u_{1s}  + u_{20} u_{1i}^* 
        &=\frac{\partial H_4}{\partial u_{1s}^*}\\
		 -i\dot{u}_{1i} &= \frac{\beta_1 \Omega^2}{2} u_{1i} + u_{20} u_{1s}^* 
         &=\frac{\partial H_4}{\partial u_{1i}^*}\\
		 -i\dot{u}_{20} &= \delta \!k u_{20} + \frac{u_{10}^2}{2} + u_{1s} u_{1i}
         &=\frac{\partial H_4}{\partial u_{20}^*}.\\
	\end{aligned}	
    \label{eq:4wm}
\end{equation} 
where the Hamiltonian is now
\begin{equation}
\begin{aligned}
H_4 =& H_4^{(0)} + H_4^{(1)}, \mathrm{with} 
\\
H_4^{(0)} &= \frac{\beta_1 \Omega^2}{2} \sum_{j=s,i} |u_{1j}|^2  +\delta k  |u_{20}|^2   \\
H_4^{(1)} &= \frac{1}{2}u_{20} (u_{10}^*)^2 + \left(u_{20} u_{1i}^* u_{1s}^* + \mathrm{c.c.}\right).
\label{eq:H4waves} 
\end{aligned}
\end{equation}

In App.~\ref{app:4waves}, we show that by exploiting physical conservation laws, we can reduce the Hamiltonian in Eq.~\eqref{eq:H4waves} to a two-degree-of-freedom system in real variables. {Yet, this does not guarantee that the MI process exhibits a regular recurrence, nor is it sufficient to predict the recurrence period as in the integrable 3WM truncation. To this end, we proceed by following a different approach.}

First, we compare in Fig.~\ref{fig10} the dynamics obtained from the PDEs Eqs.~\eqref{eq:SHGsystem} (thick solid lines) and the truncated ODEs Eqs.~\eqref{eq:4wm} (thin lines with crosses), 
{using the same perturbation parameters of Fig.~\ref{fig5}.} 
Apart from $|u_ {2s}|^2=|u_{2i}|^2$ (purple dotted line), which are not included in the truncated model and turn out, as expected, to stay very small throughout the evolution, the amplitude of oscillations of the other variables $|u_{20}|^2$, $|u_{10}|^2$ and $|u_{1s}|^2=|u_{1i}|^2$ are remarkably similar and exhibit almost regular and periodic conversion from the SH (green solid  lines) to the FF sidebands (orange dashed lines). 
The period of evolution of Eqs.~\eqref{eq:4wm} is shorter than the one extracted from Eqs.~\eqref{eq:SHGsystem}. 
{Notably, while the truncated model in Eqs.~\eqref{eq:4wm} exhibits a perfectly recurring behavior, the recurrence distance exhibited by Eqs.~\eqref{eq:SHGsystem} tends to slightly increase from one event to the next, at variance with the MI of the SH mode at larger $\delta k$ shown in Fig.~\ref{fig5}.}

Another important difference is the behavior of $|u_{10}|^2$ (blue dash-dotted lines): while the PDEs Eqs.~\eqref{eq:SHGsystem} yield an irregular behavior with multiple oscillations breaking the periodicity, Eqs.~\eqref{eq:4wm} give a more regular result, as just a single dip every period appears and the solution quickly regains its initial value.  

{ The irregular behavior of $|u_{10}|^2$ observed in the PDE evolution is mirrored by the behavior of $|u_ {2s,2i}|^2$. A more detailed analysis is needed to understand what is the role of those small sidebands in breaking the regular motion predicted by Eqs. \eqref{eq:4wm} and will be the subject of future studies.

Importantly, however, both the PDE and ODE models show that the modal intensity $|u_{10}|^2$ (as well as its conjugate phase, not shown) perform only small oscillations compared with the other variables.   }

This hints at a further simplification, based on Lie transforms perturbation theory, which yields a system identical to Eq.~\eqref{eq:Hred}, expressed in new canonical variables $(\bar\eta_{20},\bar\psi)$ obtained by the successive canonical transformations shown in App.~\ref{app:lie4waves},
\begin{equation} \label{eq:H4red}
    \bar H_{r}(\bar\eta_{20},\bar\psi)=\delta k_{4} \bar\eta_{20} + 2\sqrt{\bar\eta_{20}} (1- \bar\eta_{20}) \cos\bar\psi,
\end{equation}
with $\delta k_4 = \delta k_3 + \frac{\bar\eta_{10}}{\delta k}$, where $\bar\eta_{10}\equiv |v_{10}|^2$ is the intensity of the canonically transformed counterpart of $u_{10}$ and, most importantly, is a conserved quantity of the averaged Hamiltonian, which is thus integrable.
This result provides a justification for the regular behavior of Eqs.~\eqref{eq:4wm} shown in Fig.~\ref{fig10}. The period is thus identical, mutatis mutandis, to the result of Eq.~\eqref{eq:period}. We remark also that the difference $\delta k_4-\delta k_3$ is usually very small apart from trajectories very close to separatrix of the integrable 3WM limit.

\begin{figure}[ht!]
\begin{center}
\includegraphics[width=8cm]{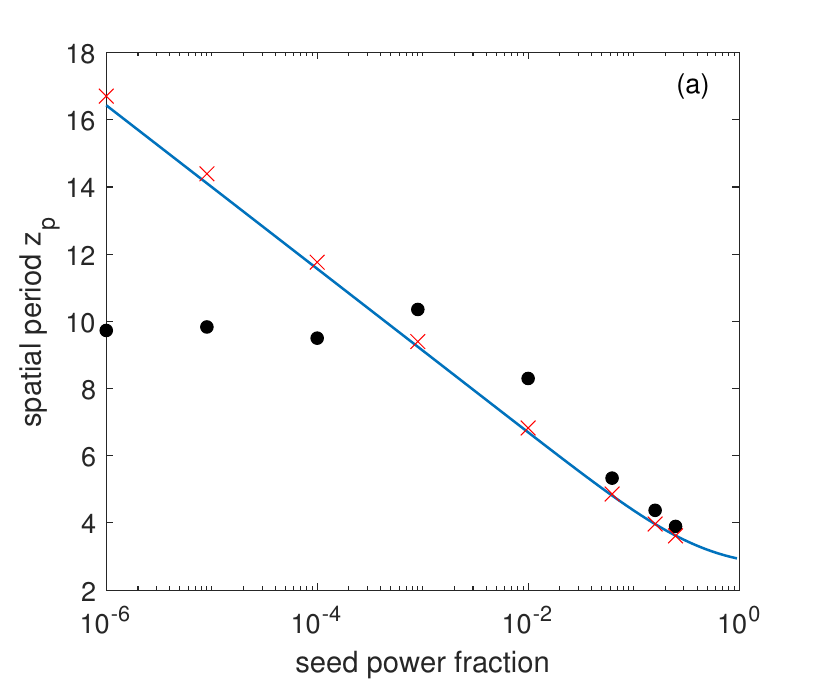}
\includegraphics[width=8cm]{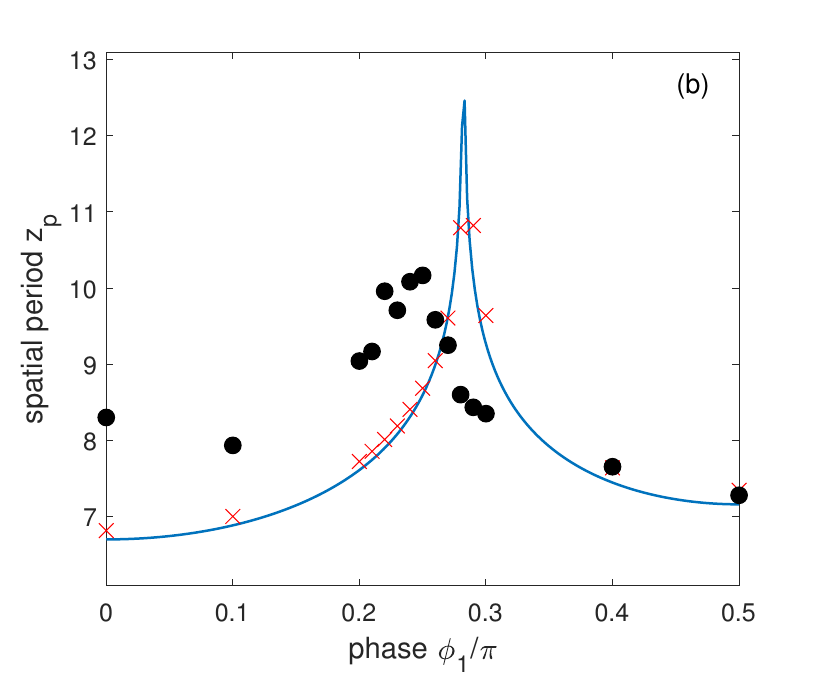}
\caption{Period of MI recurrence comparing the results from full PDE simulations of Eqs.~\eqref{eq:SHGsystem} (dots) and truncated 4WM of Eqs.~\eqref{eq:4wm} (red crosses) with the analytical estimate from Eq.~\eqref{eq:period} obtained from averaged 4WM truncation of Eq.~\eqref{eq:H4red} (solid line): (a) period vs.~power fraction of the seed for fixed phase $\phi_1=0$, $\delta k=1.8$, $\Omega_{p}=1.521$; (b) period vs.~phase of seed $\phi_1$ for fixed $c_1=0.1$.}
\label{fig11}
 \end{center}
\end{figure}

Similarly to Fig.~\ref{fig6}, in Fig.~\ref{fig11}, we compare  the period estimates (blue solid lines) as a function of $a_1^2$ of $\phi_1$ with the the recurrence period observed in simulations of Eqs.~\eqref{eq:SHGsystem} (black filled circles) and Eqs.~\eqref{eq:4wm} (red crosses). As explained above, Eqs.~\eqref{eq:SHGsystem} do not exhibit a perfectly periodic behavior; moreover, the unavoidable noise introduced by rounding errors lead to the growth of spectral components, due to spontaneous MI. Thus, the period is computed by averaging over four to six recurrences, before the noise destroys every regularity. 

In Fig.~\ref{fig11}, we first notice that the period estimates obtained from the averaged Hamiltonian are a good approximation, but generally underestimate what is found from Eqs.~\eqref{eq:4wm}. Second, similar to Fig.~\ref{fig6}, the period is very sensitive to the initial relative phases, Fig.~\ref{fig11}(b). Third, we notice that Eqs.~\eqref{eq:SHGsystem} exhibits longer periods than the analytical estimate for $a_1>0.03$. Below this point, the period saturates to a value of $z_p\approx 10$. This is a sign of a pattern of recurrence less and less regular and can be ascribed to the nonintegrability of Eqs.~\eqref{eq:SHGsystem}: presumably, for small $a_1$ the system is close to a homoclinic orbit and irregularly crosses it. The same irregular behavior is even more apparent in Fig.~\ref{fig11}(b). First we observe that the cusp predicted by Eq.~\eqref{eq:period} is at $\phi_1 = 0.28\pi$, while the period values extracted from numerical simulations reaches its largest values in the range $[0.24,0.26]$. It is difficult to observe longer periods, because often the first focusing event is followed by a slowing down of the dynamics, associated to separatrix crossing, to end up with a sequence of recurrence events with shorter periods.

\begin{figure}[ht]
\begin{center}
\includegraphics[width=.5\textwidth]{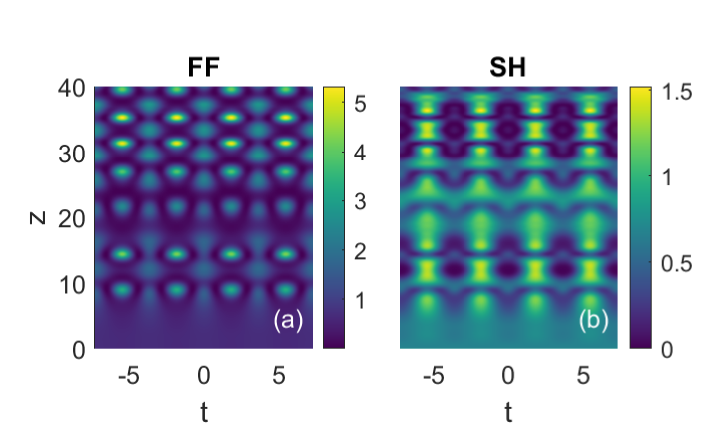}
\caption{Same as Fig.~\ref{fig9}(a-b), but for $\delta k=1.25$, $\Omega = 1.73$. The other parameters are the same as in Fig.~\ref{fig9}. The intensity profiles at the first recurrence look very similar to Fig.~\ref{fig9}(c-f) and are not reported. Notice the irregular evolution in $z$.
}
\label{fig12}
 \end{center}
\end{figure}
Finally, we show in Fig.~\ref{fig12} the spatio-temporal evolution of FF and SH for a smaller $\delta k =1.25$, at $\Omega=\Omega_p = 1.73$. The evolution is similar to Fig.~\ref{fig9} in the sense that the MI saturates and forms pulse trains in the FF. Nevertheless, it is almost impossible to identify a periodic behavior after the first focusing event. Particularly, a decomposition  similar to Fig.~\ref{fig10} shows that $|u_{10}^2|$ (not shown) starts oscillating in a very strong and irregular way, breaking the assumptions of the averaged analysis. This shows that {moving closer to the phase-matching condition of SHG, the dynamics becomes more and more complicated, and the down-conversion driven recurrence is finally lost.} 

To summarize this section, the full PDE dynamics, being non-integrable, yields a complicated behavior, that is qualitatively well described by a four-mode truncation, provided that the operating regime is not too far from the bifurcation point $\delta k = 2$. {The four-mode truncation lends itself to a further reduction to an integrable three-mode system via a suitable averaging procedure.}

\section{Conclusions} \label{s4}

In summary, we have shown that SHG admits a regime of recurrent MI dominated by down-conversion. At sufficiently large mismatch of SHG, the dynamics is well described by a three-wave truncation which is fully integrable and provides a good quantitative description of the recurrence dynamics. The regularity of the recurrence pumped by the SH starts to break down beyond the point where the SH bifurcates with a phase-locked mixed-mode eigenmode. At values of the SHG mismatch below the bifurcation point, a higher degree of regularity in the recurrence is exhibited by the MI of the phase-locked eigenmode. In this case, even the simplest truncation is no longer exactly integrable due to the presence of the fundamental component of the pump, but an accurate estimate of the recurrence period can be derived by a suitable averaging procedure. Our results provide valuable guidelines to design new experiments to  demonstrate recurrence in the MI of SHG in a regime which is somehow opposite to the cascading regime investigated to date \cite{Schiek19,Schiek21,Trillo23} { and turn out to occur on a much shorter spatial scale}. {We also point out that, such recurrence phenomena are not spoiled by the presence of a moderate group-velocity mismatch, though future work is needed to assess its specific impact.}

Due to renewed interest in quadratic media, our results will pave the way to the control and tailoring of pulse sources based on nonlinear optical waveguides.

\section*{Funding Information}  
This work was supported by European Union--Next Generation EU under funding call Progetti di Ricerca di Interesse Nazionale (PRIN), Mission 4, Component C2 (Project No. 20222NCTCY) and under the Italian National Recovery and Resilience Plan
(NRRP), Mission 4, Component 2, Investment 1.3, CUP B53C22003970001,
partnership on “Telecommunications of the Future” (PE00000001 - program
“RESTART”) - Project ODEONS.

\section*{Disclosures} The authors declare no conflicts of interest.

\section*{Data availability statement}
Data underlying the results presented in this paper are not publicly available at this time but may be obtained from the authors upon reasonable request.

\appendix
\section{The three-wave truncated model}
\label{app:3waves}

Equations (\ref{eq:3wm}) are known to be integrable \cite{Armstrong62}. Here, we write the solution by exploiting a reduction to a one degree-of-freedom Hamiltonian system, 
which generalizes the approach of Ref. \cite{Trillo92a}. To this end, we notice that the following quantities are invariants of the dynamics
\begin{eqnarray} 
|u_{20}(z)|^2 + \frac{|u_{1s}(z)|^2 + |u_{1i}(z)|^2}{2}=1; \label{eq:inv3wma} \\
|u_{1s}(z)|^2 - |u_{1i}(z)|^2 \equiv \alpha,\label{eq:inv3wmb}
\end{eqnarray}
which correspond to conservation of total intensity and Manley-Rowe (photon number difference in the sidebands), respectively.
These two invariants allow for expressing the square modulus of one variable as a function of the other two. Furthermore, it is immediately clear that only a single combination of the three phases of the fields is effective. As a consequence, the system is reducible from six to two real conjugated variables, i.e., to a single degree of freedom Hamiltonian system which is integrable by quadrature. This is explicitly done by posing $u_j=\sqrt{\eta_j} \exp(i \phi_j)$, $j=20,1s,1i$, a canonical transformation where $\eta$ and $\phi$ represent square amplitudes and phases respectively.
We obtain the following reduced equations, after simple but cumbersome algebra, by expressing Eqs.~\eqref{eq:3wm} in terms of square moduli and phases, by exploiting the invariants in Eqs. (\ref{eq:inv3wma}-\ref{eq:inv3wmb}), and by defining $\psi(z)=\phi_{20}(z)-\phi_{1s}(z)-\phi_{1i}(z)$:
\begin{eqnarray}
& \dot\eta_{20} = - \pder{H_\mathrm{3r}}{\psi};\; \dot\psi= \pder{H_\mathrm{3r}}{\eta_{20}}; \label{eq:H3wa}\\
&H_{3r} = \delta k_{3} \eta_{20} + 2 \sqrt {\eta_{20} \left[ (1-\eta_{20})^2 -\left(\frac{\alpha}{2}\right)^2 \right] } \cos{\psi},  \label{eq:H3wb}
\end{eqnarray}
where $H_{3r}$ is the reduced Hamiltonian which depends on the single parameter $\delta k_{3}$. In the case of symmetric excitation of sidebands $u_{1s}=u_{1i}$, we have $\alpha=0$, and the reduced Hamiltonian simplifies to
\begin{equation}
H_\mathrm{3r} = \delta k_{3} \eta_{20} + 2 \sqrt {\eta_{20}} (1-\eta_{20}) \cos{\psi}.  \label{eq:H3wsym}
\end{equation}
which is identical to the reduced Hamiltonian that governs SHG \cite{Trillo92a}. 
In both the symmetric and non-symmetric cases, the first of Eqs.~\eqref{eq:H3wa} can be reduced to a single equation in $\eta_{20}$ by eliminating $\psi$ through the conservation of the reduced Hamiltonian $H_{3r}$. 
For the symmetric case treated in this paper, we obtain
\begin{equation} \label{eq:eta20}
\dot\eta_{20} = 2 \sqrt{F(\eta_{20})} 
\end{equation}
where $F(\eta_{20})= \eta_{20}^3 - (2+\delta k_3^2/4)\eta_{20}^2 + (1+\delta k_3 H_{3r}/2)\eta_{20} - H_{3r}^2/4 = (\eta_{20}-a)(\eta_{20}-b)(\eta_{20}-c)$ is a third order polynomial with ordered roots $c \le b \le a$, fixed by the value of the parameter $\delta k_3$ as well as the initial condition through the value of $H_{3r}=H_{3r}(\eta_{20}(0),\psi(0))$. 
{
The down-conversion solution such that $c \le \eta_{20}(z) \le b$ and the initial value $\eta_{20}(0) = b$ can be obtained by integrating by quadrature Eq.~(\ref{eq:eta20}) in the form $\eta_{20}=c + (b-c) \sn^2(\sqrt{a-c}~(z-z_p/2) |~k)$, where the argument of the Jacobian sine $\sn$ is shifted in $z$ by half of the period $z_p$ (explicitly reported in Eq.~(\ref{eq:period})). We point out that the same solution with no-shift describes instead up-conversion occurring from the initial value $\eta_{20}(0)=c$ \cite{Armstrong62}. The down-conversion solution can also equivalently written without any shift in the fully equivalent form explicitly reported in Eq.~\eqref{eq:solution} of the main text.
}

\section{Four-wave reduction}
\label{app:4waves}
The four-mode Eqs. (\ref{eq:4wm}) are no longer integrable. Indeed, they only conserve
the Manley-Rowe relation (\ref{eq:inv3wmb}) and the total intensity which now reads
\begin{equation}    
|u_{20}(z)|^2 + \frac{|u_{10}|^2 + |u_{1s}(z)|^2 + |u_{1i}(z)|^2}{2}=1.
\label{eq:inv4waves}
\end{equation}

These two invariants allow us to express the square moduli of two of the four variables as a function of the other two. Different choices are possible yielding correspondingly different combinations of phases.
In any event, the system is reducible from eight to four real conjugated variables, i.e., to a two degrees of freedom Hamiltonian system. To this end, analogously to App.~\ref{app:3waves},  we first perform the  transformation $u_j=\sqrt{\eta_j} \exp(i \phi_j)$, with $j=10,20,1s,1i$ to obtain a real variable Hamiltonian. 
At variance with App.~\ref{app:3waves}, we have to choose two square amplitudes, instead of one. Three choices are possible and each gives different canonically conjugate phases. For instance, 
\begin{equation}
\begin{gathered}
       H_{4r} = {\beta_1 \Omega^2} \eta_{1s}  -\delta\!k \left(\frac{\eta_{10}}{2}+\eta_{1s}\right) 
       \\
       + {\eta_{10}}\sqrt{1 + \frac{\alpha}{2} - \frac{\eta_{10}}{2}-\eta_{1s}}\cos{2\psi_0}
       \\
       +2 \sqrt{\eta_{1s}(\eta_{1s} - \alpha)\left(1+ \frac{\alpha}{2}- \frac{\eta_{10}}{2}-\eta_{1s}\right)}\cos{\psi_1},
\end{gathered}
    \label{eq:Hnosidebandsred2}
\end{equation}
{
where we have eliminated $\eta_{20}$, and $\psi_1(z) \equiv  \phi_{1s}(z) + \phi_{1a}(z) - \phi_{20}(z) = -\psi(z)$ stands for the phase associated with the main down-conversion or 3WM process, whereas $\psi_0(z) \equiv \phi_{10}(z)-\frac{\phi_{20}(z)}{2}$ is the effective phase associated with the two-color pump (analogous to that used to describe SHG \cite{Trillo92a}). It is easy to verify that
}
\begin{equation}
    \begin{aligned}   
    \dot\eta_{10} &= -\pder{H_\mathrm{4r}}{\psi_0};& \dot\psi_0 = \pder{H_\mathrm{4r}}{\eta_{10}};\\
    \dot\eta_{1s} &= -\pder{H_\mathrm{4r}}{\psi_1}; & \dot\psi_1 = \pder{H_\mathrm{4r}}{\eta_{1s}},
     \end{aligned}
\end{equation}
which means that $(\eta_{10},\psi_0)$ and $(\eta_{1s},\psi_1)$ are two pairs of canonically conjugate variables. This variable choice is particularly interesting, because it reveals that the degree of freedom $(\eta_{10},\psi_0)$ exhibits only small oscillations around the SHG pump eigenmode. This hints at a the possibility of further reducing the system, by means of an averaging procedure, as illustrated in Appendix C.

\section{Lie transform averaging of four-wave system}
\label{app:lie4waves}

In Hamiltonian mechanics it is customary to define the Poisson bracket for conjugate complex variables $(u_k,u_k^*)$ as
$\{f,g\}\equiv i \left( \pder{f}{u_k}\pder{g}{u_k^*}-\pder{f}{u_k^*}\pder{g}{u_k}\right)$. This allows for expressing Eqs.~\eqref{eq:4wm} as $\dot u_k = \{u_k,H_4\}$, with $k=20,10,1s,1i$.

The averaging procedure based on Lie transform introduces a near-identity transform to new complex variables $v_k, v_k^*$ in the form $u_k = v_k + \{S, v_k\} + \{S, v_k\} + \frac{1}{2}\{S, \{S, v_k\}\} + \ldots$, where the generating function is expanded as $S = \Eps S_1 + \Eps^2 S_2 + \order {\Eps^3}$ and has to be chosen to derive an averaged Hamiltonian $\bar H = \bar H^{(0)} +\Eps \bar H^{(1)} + \Eps^2\bar H^{(2)} + \order{\Eps^3}$ of the simplest possible form \cite{Mersman1970,CaryReview1981,VerhulstBook}. The smallness parameter $\Eps$ is used here to distinguish among the different order approximations
and will be put to 1 at the end of the calculation. Notice that the $m-$th order averaged Hamiltonian $\bar H^{(m)}(v_k,v_k^*)$ is a function of the new variables.

Consider again Eqs.~\eqref{eq:H4waves}. We rewrite $H_4 = H_4^{(0)} + \Eps H_4^{(1)}$, i.e., we assume that the nonlinear terms are smaller than the linear one, which gives the unperturbed frequencies  $\omega_{10} = 0$, $\omega_{20}=\delta k$, $\omega_{1s} = \omega_{1i} =\frac{\beta_1\Omega^2}{2}$, associated to the four canonical variables.

The Lie transform approach proceeds as follows. {At $0$-th order} $\bar H_4^{(0)} = H_4^{(0)}$, whereas at order $\Eps$, we write
\begin{equation}
    \bar H_4 = H_4^{(1)} + \{S_1,H_4^{(0)}\}.
\end{equation}
We notice that $H_4^{(1)}$ includes two  monomials, $\mathcal{M}_1\equiv \frac{1}{2}u_{20} (u_{10}^*)^2$, which oscillate at frequencies $\Delta\omega_1 =  \delta k$ and $\mathcal{M}_2 \equiv u_{20} u_{1i}^* u_{1s}^* $, which oscillate at frequencies $\Delta\omega_2 = \omega_{20} - \omega_{1s}-\omega_{1i} =  \delta k - \beta_1\Omega^2$, as long as their complex conjugates $\mathcal{M}_{p}^*$ associated to $-\Delta\omega_{p}$, with $p=1,2$. As shown in Sec.~\ref{s2}, the conversion is most effective when $\delta k_3 \approx 0$ 
{(striclty, $\delta k_3 = 0$ only at the bifurcation point $\delta k=2$). Therefore,} we conclude that $\Delta\omega_2\approx 0$ and the associated terms in $H_4^{(1)}$ have to be considered as resonant; the generating function $S_1$ is chosen to eliminate (i.e., average out) the non-resonant terms, associated to $\pm\Delta\omega_1$. The construction of the generating function is straightforward: $S_m = i\sum_p{\frac{\mathcal{M}_p}{\Delta\omega_p}}$, where the sum is performed over non-resonant monomials (including their complex conjugates).
We obtain
\begin{equation}
    \bar H_4^{(1)} = v_{20} v_{1i}^* v_{1s}^* + \mathrm{c.c.}, \; S_1 =  \frac{i}{2\delta k} (v_{20} (v_{10}^*)^2 - \mathrm{c.c.}).
    \label{eq:Hbar1_S1}
\end{equation}
We notice that, up to first order, the averaged Hamiltonian $\bar H_4^{(0)}+\bar H_4^{(1)}$ is identical to the $H_3$ of Eq.~\eqref{eq:H3}.

At order ${\Eps^2}$, the Lie series gives
\begin{equation}
    \bar H_4^{(2)} = \{S_2,H_4^{(0)}\} + \{S_1, H_4^{(1)}\} + \frac{1}{2}\{S_1, \{S_1, H_4^{(0)}\}\}.
\end{equation}
Analyzing each term appearing in this expression, we eliminate the non-resonant ones by a suitable choice of $S_2$. We obtain
\begin{equation}
\begin{gathered}
    \bar H_4^{(2)} = -\frac{1}{4\delta k} (|v_{10}|^4 - 4|v_{10}|^2 |v_{20}|^2),
    \\
    S_2 = -\frac{i}{4\Delta\delta k} ((v_{10}^*)^2 v_{1s} v_{1i} - v_{10}^2 v_{1s}^* v_{1i}^*)
    \end{gathered}
     \label{eq:Hbar2_S2}
\end{equation}

First notice that $v_{10}$ appears only in $\bar H_4^{(2)}$ through $\bar \eta_{10}\equiv |v_{10}|^2$, which  is thus a conserved quantity of the evolution. It is also easy to verify that the following quantities are invariants of the dynamics, because they are in involution with $\bar H_1$
\begin{equation}
    \begin{aligned}
|v_{20}(z)|^2 + \frac{|v_{1s}(z)|^2 + |v_{1i}(z)|^2}{2} = \Lambda;  \\
|v_{1s}(z)|^2 - |v_{1i}(z)|^2 \equiv \alpha.
\label{eq:invave4waves}
    \end{aligned}
\end{equation}
For $\bar H_4$ has four conserved quantities in involution, it turns out to be integrable to second order. 

By transforming to real variables $v_j = \sqrt{\bar\eta_j}e^{i\phi_j}$, neglecting ineffective constant terms, and eliminating $\bar\eta_{1s}$ and $\bar\eta_{1i}$ via Eq.~\eqref{eq:invave4waves}, we obtain
\begin{eqnarray}
& \dot{\bar{\eta}}_{20} = - \pder{\bar H_\mathrm{4r}}{\bar\psi};\; \dot{\bar\psi} = \pder{\bar H_\mathrm{4r}}{\bar \eta_{20}}; \label{eq:H4wa}\\
&\bar H_{4r} = \delta k_{4} \bar\eta_{20} + 2 \sqrt {\bar \eta_{20} \left[ (1-\bar \eta_{20})^2 -\left(\frac{\alpha}{2}\right)^2 \right] } \cos{\bar \psi},  \label{eq:H4wb}
\end{eqnarray}
{
with an effective mismatch $\delta k_4 = \delta k_3 + \frac{\bar\eta_{10}}{\delta k}$. For the symmetric case $\alpha=0$,
we obtain Eq.~\eqref{eq:H4red} of the main text. Eqs.~\eqref{eq:H4wa} are identical to Eqs.~\eqref{eq:H3wa} and their integration by quadrature is performed as explained in App.~\ref{app:3waves}.
} 
\newpage
For the sake of completeness, we explicitly write the canonical transformations, which read
\begin{equation}
\begin{aligned}
v_{10} &= u_{10} - \frac{1}{\delta k} u_{10}^* u_{20} + \frac{1}{2\beta_1\Omega^2 \delta k} u_{10}^* u_{1s} u_{1i},\\
&+ \frac{u_{10}}{4\delta k^2} (2|u_{20}|^2 - |u_{10}|^2), \\
v_{20} &= u_{20} + \frac{1}{2\delta k} u_{10}^2 - \frac{1}{2\delta k^2} |u_{10}|^2 u_{20}, \\
v_{1s} &= u_{1s} - \frac{1}{4\beta_1\Omega^2\delta k} u_{10}^2 u_{1i}^*, \\
v_{1i} &= u_{1i} - \frac{1}{4\beta_1\Omega^2\delta k} u_{10}^2 u_{1s}^*.
\end{aligned}
\label{eq:inversecanonical}
\end{equation}

\newpage
The inverse transformation to original variables read
\begin{equation}
\begin{aligned}
u_{10} &= v_{10} + \frac{1}{\delta k} v_{10}^* v_{20} - \frac{1}{2\beta_1\Omega^2 \delta k} v_{10}^* v_{1s} v_{1i}, \\
&+ \frac{v_{10}}{4\delta k^2} (2|v_{20}|^2 - |v_{10}|^2), \\
u_{20} &= v_{20} - \frac{1}{2\delta k} v_{10}^2 - \frac{1}{2\delta k^2} |v_{10}|^2 v_{20}, \\
u_{1s} &= v_{1s} + \frac{1}{4\beta_1\Omega^2\delta k} v_{10}^2 v_{1i}^*, \\
u_{1i} &= v_{1i} + \frac{1}{4\beta_1\Omega^2\delta k} v_{10}^2 v_{1s}^*.
\end{aligned}
\label{eq:directcanonical}
\end{equation}

The calculation of approximate closed form solutions of Eqs.~\eqref{eq:4wm} can be done by transforming initial conditions to the new variables by means of Eqs.~\eqref{eq:inversecanonical}, solving Eq.~\eqref{eq:H4wa} as illustrated in App.~\ref{app:3waves} and transforming back to original variables by means of Eqs.~\eqref{eq:directcanonical}. In the main text we limit ourselves to the calculation of the period of oscillations, which does not require the last step.


\begin{thebibliography}{42}%
\makeatletter
\providecommand \@ifxundefined [1]{%
 \@ifx{#1\undefined}
}%
\providecommand \@ifnum [1]{%
 \ifnum #1\expandafter \@firstoftwo
 \else \expandafter \@secondoftwo
 \fi
}%
\providecommand \@ifx [1]{%
 \ifx #1\expandafter \@firstoftwo
 \else \expandafter \@secondoftwo
 \fi
}%
\providecommand \natexlab [1]{#1}%
\providecommand \enquote  [1]{``#1''}%
\providecommand \bibnamefont  [1]{#1}%
\providecommand \bibfnamefont [1]{#1}%
\providecommand \citenamefont [1]{#1}%
\providecommand \href@noop [0]{\@secondoftwo}%
\providecommand \href [0]{\begingroup \@sanitize@url \@href}%
\providecommand \@href[1]{\@@startlink{#1}\@@href}%
\providecommand \@@href[1]{\endgroup#1\@@endlink}%
\providecommand \@sanitize@url [0]{\catcode `\\12\catcode `\$12\catcode `\&12\catcode `\#12\catcode `\^12\catcode `\_12\catcode `\%12\relax}%
\providecommand \@@startlink[1]{}%
\providecommand \@@endlink[0]{}%
\providecommand \url  [0]{\begingroup\@sanitize@url \@url }%
\providecommand \@url [1]{\endgroup\@href {#1}{\urlprefix }}%
\providecommand \urlprefix  [0]{URL }%
\providecommand \Eprint [0]{\href }%
\providecommand \doibase [0]{https://doi.org/}%
\providecommand \selectlanguage [0]{\@gobble}%
\providecommand \bibinfo  [0]{\@secondoftwo}%
\providecommand \bibfield  [0]{\@secondoftwo}%
\providecommand \translation [1]{[#1]}%
\providecommand \BibitemOpen [0]{}%
\providecommand \bibitemStop [0]{}%
\providecommand \bibitemNoStop [0]{.\EOS\space}%
\providecommand \EOS [0]{\spacefactor3000\relax}%
\providecommand \BibitemShut  [1]{\csname bibitem#1\endcsname}%
\let\auto@bib@innerbib\@empty
\bibitem [{\citenamefont {Bespalov}\ and\ \citenamefont {Talanov}(1966)}]{BT}%
  \BibitemOpen
  \bibfield  {author} {\bibinfo {author} {\bibfnamefont {V.~I.}\ \bibnamefont {Bespalov}}\ and\ \bibinfo {author} {\bibfnamefont {V.~I.}\ \bibnamefont {Talanov}},\ }\bibfield  {title} {\bibinfo {title} {{Filamentary structure of light beams in nonlinear liquids}},\ }\href@noop {} {\bibfield  {journal} {\bibinfo  {journal} {Sov. Phys. JETP Lett.}\ }\textbf {\bibinfo {volume} {3}},\ \bibinfo {pages} {307} (\bibinfo {year} {1966})}\BibitemShut {NoStop}%
\bibitem [{\citenamefont {Tai}\ \emph {et~al.}(1986)\citenamefont {Tai}, \citenamefont {Hasegawa},\ and\ \citenamefont {Tomita}}]{Tai86}%
  \BibitemOpen
  \bibfield  {author} {\bibinfo {author} {\bibfnamefont {K.}~\bibnamefont {Tai}}, \bibinfo {author} {\bibfnamefont {A.}~\bibnamefont {Hasegawa}},\ and\ \bibinfo {author} {\bibfnamefont {A.}~\bibnamefont {Tomita}},\ }\bibfield  {title} {\bibinfo {title} {{Observation of modulational instability in optical fibers}},\ }\href@noop {} {\bibfield  {journal} {\bibinfo  {journal} {Phys. Rev. Lett.}\ }\textbf {\bibinfo {volume} {56}},\ \bibinfo {pages} {135} (\bibinfo {year} {1986})}\BibitemShut {NoStop}%
\bibitem [{\citenamefont {Van~Simaeys}\ \emph {et~al.}(2001)\citenamefont {Van~Simaeys}, \citenamefont {Emplit},\ and\ \citenamefont {Haelterman}}]{VanSim01}%
  \BibitemOpen
  \bibfield  {author} {\bibinfo {author} {\bibfnamefont {G.}~\bibnamefont {Van~Simaeys}}, \bibinfo {author} {\bibfnamefont {P.}~\bibnamefont {Emplit}},\ and\ \bibinfo {author} {\bibfnamefont {M.}~\bibnamefont {Haelterman}},\ }\bibfield  {title} {\bibinfo {title} {{Experimental demonstration of the Fermi-Pasta-Ulam recurrence in a modulationally unstable optical wave}},\ }\href@noop {} {\bibfield  {journal} {\bibinfo  {journal} {Phys. Rev. Lett.}\ }\textbf {\bibinfo {volume} {87}},\ \bibinfo {pages} {033902} (\bibinfo {year} {2001})}\BibitemShut {NoStop}%
\bibitem [{\citenamefont {Mussot}\ \emph {et~al.}(2018)\citenamefont {Mussot}, \citenamefont {Naveau}, \citenamefont {Conforti}, \citenamefont {Kudlinski}, \citenamefont {Copie}, \citenamefont {Szriftgiser},\ and\ \citenamefont {Trillo}}]{Mussot18}%
  \BibitemOpen
  \bibfield  {author} {\bibinfo {author} {\bibfnamefont {A.}~\bibnamefont {Mussot}}, \bibinfo {author} {\bibfnamefont {C.}~\bibnamefont {Naveau}}, \bibinfo {author} {\bibfnamefont {M.}~\bibnamefont {Conforti}}, \bibinfo {author} {\bibfnamefont {A.}~\bibnamefont {Kudlinski}}, \bibinfo {author} {\bibfnamefont {F.}~\bibnamefont {Copie}}, \bibinfo {author} {\bibfnamefont {P.}~\bibnamefont {Szriftgiser}},\ and\ \bibinfo {author} {\bibfnamefont {S.}~\bibnamefont {Trillo}},\ }\bibfield  {title} {\bibinfo {title} {{Fibre multi-wave mixing combs reveal the broken symmetry of Fermi-Pasta-Ulam recurrence}},\ }\href@noop {} {\bibfield  {journal} {\bibinfo  {journal} {Nat. Photonics}\ }\textbf {\bibinfo {volume} {12}},\ \bibinfo {pages} {303} (\bibinfo {year} {2018})}\BibitemShut {NoStop}%
\bibitem [{\citenamefont {Pierangeli}\ \emph {et~al.}(2018)\citenamefont {Pierangeli}, \citenamefont {Flammini}, \citenamefont {Zhang}, \citenamefont {Marcucci}, \citenamefont {Agranat}, \citenamefont {Grinevich}, \citenamefont {Santini}, \citenamefont {Conti},\ and\ \citenamefont {Del~Re}}]{Pierangeli18}%
  \BibitemOpen
  \bibfield  {author} {\bibinfo {author} {\bibfnamefont {D.}~\bibnamefont {Pierangeli}}, \bibinfo {author} {\bibfnamefont {M.}~\bibnamefont {Flammini}}, \bibinfo {author} {\bibfnamefont {L.}~\bibnamefont {Zhang}}, \bibinfo {author} {\bibfnamefont {G.}~\bibnamefont {Marcucci}}, \bibinfo {author} {\bibfnamefont {A.}~\bibnamefont {Agranat}}, \bibinfo {author} {\bibfnamefont {P.~G.}\ \bibnamefont {Grinevich}}, \bibinfo {author} {\bibfnamefont {P.~M.}\ \bibnamefont {Santini}}, \bibinfo {author} {\bibfnamefont {C.}~\bibnamefont {Conti}},\ and\ \bibinfo {author} {\bibfnamefont {E.}~\bibnamefont {Del~Re}},\ }\bibfield  {title} {\bibinfo {title} {{Observation of Fermi-Pasta-Ulam-Tsingou Recurrence and Its Exact Dynamics}},\ }\href@noop {} {\bibfield  {journal} {\bibinfo  {journal} {Phys. Rev. X}\ }\textbf {\bibinfo {volume} {8}},\ \bibinfo {pages} {041017} (\bibinfo {year} {2018})}\BibitemShut {NoStop}%
\bibitem [{\citenamefont {Goossens}\ \emph {et~al.}(2019)\citenamefont {Goossens}, \citenamefont {Hafermann},\ and\ \citenamefont {Jaoun}}]{Goossens19}%
  \BibitemOpen
  \bibfield  {author} {\bibinfo {author} {\bibfnamefont {J.-W.}\ \bibnamefont {Goossens}}, \bibinfo {author} {\bibfnamefont {H.}~\bibnamefont {Hafermann}},\ and\ \bibinfo {author} {\bibfnamefont {Y.}~\bibnamefont {Jaoun}},\ }\bibfield  {title} {\bibinfo {title} {{Experimental realization of Fermi-Pasta-Ulam-Tsingou recurrence in a long-haul optical fiber transmission system}},\ }\href@noop {} {\bibfield  {journal} {\bibinfo  {journal} {Sci. Rep.}\ }\textbf {\bibinfo {volume} {9}},\ \bibinfo {pages} {1} (\bibinfo {year} {2019})}\BibitemShut {NoStop}%
\bibitem [{\citenamefont {Vanderhaegen}\ \emph {et~al.}(2021)\citenamefont {Vanderhaegen}, \citenamefont {Naveau}, \citenamefont {Szriftgiser}, \citenamefont {Kudlinski}, \citenamefont {Conforti}, \citenamefont {Mussot}, \citenamefont {Onorato}, \citenamefont {Trillo}, \citenamefont {Chabchoub},\ and\ \citenamefont {Akhmediev}}]{Van21}%
  \BibitemOpen
  \bibfield  {author} {\bibinfo {author} {\bibfnamefont {G.}~\bibnamefont {Vanderhaegen}}, \bibinfo {author} {\bibfnamefont {C.}~\bibnamefont {Naveau}}, \bibinfo {author} {\bibfnamefont {P.}~\bibnamefont {Szriftgiser}}, \bibinfo {author} {\bibfnamefont {A.}~\bibnamefont {Kudlinski}}, \bibinfo {author} {\bibfnamefont {M.}~\bibnamefont {Conforti}}, \bibinfo {author} {\bibfnamefont {A.}~\bibnamefont {Mussot}}, \bibinfo {author} {\bibfnamefont {M.}~\bibnamefont {Onorato}}, \bibinfo {author} {\bibfnamefont {S.}~\bibnamefont {Trillo}}, \bibinfo {author} {\bibfnamefont {A.}~\bibnamefont {Chabchoub}},\ and\ \bibinfo {author} {\bibfnamefont {N.}~\bibnamefont {Akhmediev}},\ }\bibfield  {title} {\bibinfo {title} {{``Extraordinary" modulation instability in optics and hydrodynamics}},\ }\href@noop {} {\bibfield  {journal} {\bibinfo  {journal} {PNAS}\ }\textbf {\bibinfo {volume} {118}},\ \bibinfo {pages} {e2019348118} (\bibinfo {year} {2021})}\BibitemShut {NoStop}%
\bibitem [{\citenamefont {Fermi}\ \emph {et~al.}(1965)\citenamefont {Fermi}, \citenamefont {Pasta},\ and\ \citenamefont {Ulam}}]{FPUT}%
  \BibitemOpen
  \bibfield  {author} {\bibinfo {author} {\bibfnamefont {E.}~\bibnamefont {Fermi}}, \bibinfo {author} {\bibfnamefont {J.}~\bibnamefont {Pasta}},\ and\ \bibinfo {author} {\bibfnamefont {S.}~\bibnamefont {Ulam}},\ }\bibfield  {title} {\bibinfo {title} {{Collected Papers of Enrico Fermi}},\ }in\ \href@noop {} {\emph {\bibinfo {booktitle} {Collected Papers of Enrico Fermi}}},\ Vol.~\bibinfo {volume} {2},\ \bibinfo {editor} {edited by\ \bibinfo {editor} {\bibfnamefont {E.}~\bibnamefont {Segr{\'e}}}}\ (\bibinfo  {publisher} {The University of Chicago},\ \bibinfo {address} {Chicago},\ \bibinfo {year} {1965})\ pp.\ \bibinfo {pages} {977--988}\BibitemShut {NoStop}%
\bibitem [{\citenamefont {Armaroli}\ and\ \citenamefont {Trillo}(2024)}]{Armaroli24}%
  \BibitemOpen
  \bibfield  {author} {\bibinfo {author} {\bibfnamefont {A.}~\bibnamefont {Armaroli}}\ and\ \bibinfo {author} {\bibfnamefont {S.}~\bibnamefont {Trillo}},\ }\bibfield  {title} {\bibinfo {title} {{Recurrent nonlinear modulational instability in the $\beta$-FPUT chain}},\ }\href@noop {} {\bibfield  {journal} {\bibinfo  {journal} {Chaos, Solitons, \& Fractals}\ }\textbf {\bibinfo {volume} {188}},\ \bibinfo {pages} {115573} (\bibinfo {year} {2024})}\BibitemShut {NoStop}%
\bibitem [{\citenamefont {Conforti}\ \emph {et~al.}(2016)\citenamefont {Conforti}, \citenamefont {Armaroli}, \citenamefont {Kudlinski}, \citenamefont {Mussot}, \citenamefont {Rota~Nodari}, \citenamefont {Dujardin}, \citenamefont {De~Bi\`evre},\ and\ \citenamefont {Trillo}}]{Conforti16}%
  \BibitemOpen
  \bibfield  {author} {\bibinfo {author} {\bibfnamefont {M.}~\bibnamefont {Conforti}}, \bibinfo {author} {\bibfnamefont {A.}~\bibnamefont {Armaroli}}, \bibinfo {author} {\bibfnamefont {A.}~\bibnamefont {Kudlinski}}, \bibinfo {author} {\bibfnamefont {A.}~\bibnamefont {Mussot}}, \bibinfo {author} {\bibfnamefont {S.}~\bibnamefont {Rota~Nodari}}, \bibinfo {author} {\bibfnamefont {G.}~\bibnamefont {Dujardin}}, \bibinfo {author} {\bibfnamefont {S.}~\bibnamefont {De~Bi\`evre}},\ and\ \bibinfo {author} {\bibfnamefont {S.}~\bibnamefont {Trillo}},\ }\bibfield  {title} {\bibinfo {title} {{Heteroclinic structure of parametric resonance in the nonlinear Schr\"odinger equation}},\ }\href@noop {} {\bibfield  {journal} {\bibinfo  {journal} {Phys. Rev. Lett.}\ }\textbf {\bibinfo {volume} {117}},\ \bibinfo {pages} {013901} (\bibinfo {year} {2016})}\BibitemShut {NoStop}%
\bibitem [{\citenamefont {Yao}\ \emph {et~al.}(2022)\citenamefont {Yao}, \citenamefont {Liu}, \citenamefont {Yang},\ and\ \citenamefont {Yang}}]{Yao22}%
  \BibitemOpen
  \bibfield  {author} {\bibinfo {author} {\bibfnamefont {X.}~\bibnamefont {Yao}}, \bibinfo {author} {\bibfnamefont {C.}~\bibnamefont {Liu}}, \bibinfo {author} {\bibfnamefont {Z.-Y.}\ \bibnamefont {Yang}},\ and\ \bibinfo {author} {\bibfnamefont {W.-L.}\ \bibnamefont {Yang}},\ }\bibfield  {title} {\bibinfo {title} {{Heteroclinic-structure transition of the pure quartic modulation instability}},\ }\href@noop {} {\bibfield  {journal} {\bibinfo  {journal} {Phys. Rev. Res.}\ }\textbf {\bibinfo {volume} {4}},\ \bibinfo {pages} {013246} (\bibinfo {year} {2022})}\BibitemShut {NoStop}%
\bibitem [{\citenamefont {Trillo}\ and\ \citenamefont {Ferro}(1995)}]{TF95a}%
  \BibitemOpen
  \bibfield  {author} {\bibinfo {author} {\bibfnamefont {S.}~\bibnamefont {Trillo}}\ and\ \bibinfo {author} {\bibfnamefont {P.}~\bibnamefont {Ferro}},\ }\bibfield  {title} {\bibinfo {title} {{Modulational instability in second-harmonic generation}},\ }\href@noop {} {\bibfield  {journal} {\bibinfo  {journal} {Opt. Lett.}\ }\textbf {\bibinfo {volume} {20}},\ \bibinfo {pages} {438} (\bibinfo {year} {1995})}\BibitemShut {NoStop}%
\bibitem [{\citenamefont {Ferro}\ and\ \citenamefont {Trillo}(1995)}]{TF95b}%
  \BibitemOpen
  \bibfield  {author} {\bibinfo {author} {\bibfnamefont {P.}~\bibnamefont {Ferro}}\ and\ \bibinfo {author} {\bibfnamefont {S.}~\bibnamefont {Trillo}},\ }\bibfield  {title} {\bibinfo {title} {{Periodic waves, domain walls and modulational instability in dispersive quadratic media}},\ }\href@noop {} {\bibfield  {journal} {\bibinfo  {journal} {Phys. Rev. E}\ }\textbf {\bibinfo {volume} {51}},\ \bibinfo {pages} {4994} (\bibinfo {year} {1995})}\BibitemShut {NoStop}%
\bibitem [{\citenamefont {Buryak}\ and\ \citenamefont {Kivshar}(1995)}]{Buryak95}%
  \BibitemOpen
  \bibfield  {author} {\bibinfo {author} {\bibfnamefont {A.~V.}\ \bibnamefont {Buryak}}\ and\ \bibinfo {author} {\bibfnamefont {Y.~S.}\ \bibnamefont {Kivshar}},\ }\bibfield  {title} {\bibinfo {title} {{Dark solitons in dispersive quadratic media}},\ }\href@noop {} {\bibfield  {journal} {\bibinfo  {journal} {Opt. Lett.}\ }\textbf {\bibinfo {volume} {20}},\ \bibinfo {pages} {834} (\bibinfo {year} {1995})}\BibitemShut {NoStop}%
\bibitem [{\citenamefont {Kennedy}\ and\ \citenamefont {Trillo}(1996)}]{Kennedy96}%
  \BibitemOpen
  \bibfield  {author} {\bibinfo {author} {\bibfnamefont {T.~A.~B.}\ \bibnamefont {Kennedy}}\ and\ \bibinfo {author} {\bibfnamefont {S.}~\bibnamefont {Trillo}},\ }\bibfield  {title} {\bibinfo {title} {{Squeezing of cw light in a single-mode dispersive $\chi^{(2)}$ medium}},\ }\href@noop {} {\bibfield  {journal} {\bibinfo  {journal} {Phys. Rev. A}\ }\textbf {\bibinfo {volume} {54}},\ \bibinfo {pages} {4396} (\bibinfo {year} {1996})}\BibitemShut {NoStop}%
\bibitem [{\citenamefont {Trillo}\ and\ \citenamefont {Wabnitz}(1997)}]{Trillo97}%
  \BibitemOpen
  \bibfield  {author} {\bibinfo {author} {\bibfnamefont {S.}~\bibnamefont {Trillo}}\ and\ \bibinfo {author} {\bibfnamefont {S.}~\bibnamefont {Wabnitz}},\ }\bibfield  {title} {\bibinfo {title} {{Dynamic spontaneous fluorescence in parametric wave coupling}},\ }\href@noop {} {\bibfield  {journal} {\bibinfo  {journal} {Phys. Rev. E}\ }\textbf {\bibinfo {volume} {55}},\ \bibinfo {pages} {R4897} (\bibinfo {year} {1997})}\BibitemShut {NoStop}%
\bibitem [{\citenamefont {Buryak}\ \emph {et~al.}(2002)\citenamefont {Buryak}, \citenamefont {Di~Trapani}, \citenamefont {Skryabin},\ and\ \citenamefont {Trillo}}]{Buryak02}%
  \BibitemOpen
  \bibfield  {author} {\bibinfo {author} {\bibfnamefont {A.~V.}\ \bibnamefont {Buryak}}, \bibinfo {author} {\bibfnamefont {P.}~\bibnamefont {Di~Trapani}}, \bibinfo {author} {\bibfnamefont {D.}~\bibnamefont {Skryabin}},\ and\ \bibinfo {author} {\bibfnamefont {S.}~\bibnamefont {Trillo}},\ }\bibfield  {title} {\bibinfo {title} {{Optical solitons due to quadratic nonlinearities: from basic physics to futuristic applications}},\ }\href@noop {} {\bibfield  {journal} {\bibinfo  {journal} {Phys. Rep.}\ }\textbf {\bibinfo {volume} {370}},\ \bibinfo {pages} {63} (\bibinfo {year} {2002})}\BibitemShut {NoStop}%
\bibitem [{\citenamefont {Fuerst}\ \emph {et~al.}(1997)\citenamefont {Fuerst}, \citenamefont {Baboiu}, \citenamefont {Lawrence}, \citenamefont {Torruellas}, \citenamefont {Stegeman}, \citenamefont {Trillo},\ and\ \citenamefont {Wabnitz}}]{Fuerts97}%
  \BibitemOpen
  \bibfield  {author} {\bibinfo {author} {\bibfnamefont {R.~A.}\ \bibnamefont {Fuerst}}, \bibinfo {author} {\bibfnamefont {D.~M.}\ \bibnamefont {Baboiu}}, \bibinfo {author} {\bibfnamefont {B.}~\bibnamefont {Lawrence}}, \bibinfo {author} {\bibfnamefont {W.~E.}\ \bibnamefont {Torruellas}}, \bibinfo {author} {\bibfnamefont {G.~I.}\ \bibnamefont {Stegeman}}, \bibinfo {author} {\bibfnamefont {S.}~\bibnamefont {Trillo}},\ and\ \bibinfo {author} {\bibfnamefont {S.}~\bibnamefont {Wabnitz}},\ }\bibfield  {title} {\bibinfo {title} {{Spatial modulational instability and multisoliton-like generation in a quadratically nonlinear optical medium}},\ }\href@noop {} {\bibfield  {journal} {\bibinfo  {journal} {Phys. Rev. Lett.}\ }\textbf {\bibinfo {volume} {78}},\ \bibinfo {pages} {2756} (\bibinfo {year} {1997})}\BibitemShut {NoStop}%
\bibitem [{\citenamefont {Fang}\ \emph {et~al.}(2000)\citenamefont {Fang}, \citenamefont {Malendevich}, \citenamefont {Schiek},\ and\ \citenamefont {Stegeman}}]{Fang00}%
  \BibitemOpen
  \bibfield  {author} {\bibinfo {author} {\bibfnamefont {H.}~\bibnamefont {Fang}}, \bibinfo {author} {\bibfnamefont {R.}~\bibnamefont {Malendevich}}, \bibinfo {author} {\bibfnamefont {R.}~\bibnamefont {Schiek}},\ and\ \bibinfo {author} {\bibfnamefont {G.~I.}\ \bibnamefont {Stegeman}},\ }\bibfield  {title} {\bibinfo {title} {{Spatial modulational instability in one-dimensional lithium niobate slab waveguides}},\ }\href@noop {} {\bibfield  {journal} {\bibinfo  {journal} {Opt. Lett.}\ }\textbf {\bibinfo {volume} {25}},\ \bibinfo {pages} {1786} (\bibinfo {year} {2000})}\BibitemShut {NoStop}%
\bibitem [{\citenamefont {Schiek}\ \emph {et~al.}(2001)\citenamefont {Schiek}, \citenamefont {Fang}, \citenamefont {Malendevich},\ and\ \citenamefont {Stegeman}}]{Schiek01}%
  \BibitemOpen
  \bibfield  {author} {\bibinfo {author} {\bibfnamefont {R.}~\bibnamefont {Schiek}}, \bibinfo {author} {\bibfnamefont {H.}~\bibnamefont {Fang}}, \bibinfo {author} {\bibfnamefont {R.}~\bibnamefont {Malendevich}},\ and\ \bibinfo {author} {\bibfnamefont {G.~I.}\ \bibnamefont {Stegeman}},\ }\bibfield  {title} {\bibinfo {title} {{Measurement of modulational instability gain of second-order nonlinear optical eigenmodes in a one-dimensional system}},\ }\href@noop {} {\bibfield  {journal} {\bibinfo  {journal} {Phys. Rev. Lett.}\ }\textbf {\bibinfo {volume} {86}},\ \bibinfo {pages} {4528} (\bibinfo {year} {2001})}\BibitemShut {NoStop}%
\bibitem [{\citenamefont {Delqu{\'e}}\ \emph {et~al.}(2011)\citenamefont {Delqu{\'e}}, \citenamefont {Fanjoux}, \citenamefont {Gorza},\ and\ \citenamefont {Haelterman}}]{Delque11}%
  \BibitemOpen
  \bibfield  {author} {\bibinfo {author} {\bibfnamefont {M.}~\bibnamefont {Delqu{\'e}}}, \bibinfo {author} {\bibfnamefont {G.}~\bibnamefont {Fanjoux}}, \bibinfo {author} {\bibfnamefont {S.-P.}\ \bibnamefont {Gorza}},\ and\ \bibinfo {author} {\bibfnamefont {M.}~\bibnamefont {Haelterman}},\ }\bibfield  {title} {\bibinfo {title} {{Spontaneous 2D modulation instability in second harmonic generation process}},\ }\href@noop {} {\bibfield  {journal} {\bibinfo  {journal} {Opt. Commun.}\ }\textbf {\bibinfo {volume} {284}},\ \bibinfo {pages} {1401} (\bibinfo {year} {2011})}\BibitemShut {NoStop}%
\bibitem [{\citenamefont {Jauberteau}\ \emph {et~al.}(2021)\citenamefont {Jauberteau}, \citenamefont {Wehbi}, \citenamefont {Mansuryan}, \citenamefont {Krupa}, \citenamefont {Baronio}, \citenamefont {Wetzel}, \citenamefont {Tonello}, \citenamefont {Wabnitz},\ and\ \citenamefont {Couderc}}]{Jab21}%
  \BibitemOpen
  \bibfield  {author} {\bibinfo {author} {\bibfnamefont {R.}~\bibnamefont {Jauberteau}}, \bibinfo {author} {\bibfnamefont {S.}~\bibnamefont {Wehbi}}, \bibinfo {author} {\bibfnamefont {T.}~\bibnamefont {Mansuryan}}, \bibinfo {author} {\bibfnamefont {K.}~\bibnamefont {Krupa}}, \bibinfo {author} {\bibfnamefont {F.}~\bibnamefont {Baronio}}, \bibinfo {author} {\bibfnamefont {B.}~\bibnamefont {Wetzel}}, \bibinfo {author} {\bibfnamefont {A.}~\bibnamefont {Tonello}}, \bibinfo {author} {\bibfnamefont {S.}~\bibnamefont {Wabnitz}},\ and\ \bibinfo {author} {\bibfnamefont {V.}~\bibnamefont {Couderc}},\ }\bibfield  {title} {\bibinfo {title} {{Boosting and taming wave breakup in second harmonic generation}},\ }\href@noop {} {\bibfield  {journal} {\bibinfo  {journal} {Front. Phys.}\ }\textbf {\bibinfo {volume} {9}},\ \bibinfo {pages} {640025} (\bibinfo {year} {2021})}\BibitemShut {NoStop}%
\bibitem [{\citenamefont {Salerno}\ \emph {et~al.}(2003)\citenamefont {Salerno}, \citenamefont {Minardi}, \citenamefont {Trull}, \citenamefont {Varanavicius}, \citenamefont {Tamosauskas}, \citenamefont {Valiulis}, \citenamefont {Dubietis}, \citenamefont {Caironi}, \citenamefont {Trillo}, \citenamefont {Piskarskas},\ and\ \citenamefont {Di~Trapani}}]{Salerno03}%
  \BibitemOpen
  \bibfield  {author} {\bibinfo {author} {\bibfnamefont {D.}~\bibnamefont {Salerno}}, \bibinfo {author} {\bibfnamefont {S.}~\bibnamefont {Minardi}}, \bibinfo {author} {\bibfnamefont {J.}~\bibnamefont {Trull}}, \bibinfo {author} {\bibfnamefont {A.}~\bibnamefont {Varanavicius}}, \bibinfo {author} {\bibfnamefont {G.}~\bibnamefont {Tamosauskas}}, \bibinfo {author} {\bibfnamefont {G.}~\bibnamefont {Valiulis}}, \bibinfo {author} {\bibfnamefont {A.}~\bibnamefont {Dubietis}}, \bibinfo {author} {\bibfnamefont {D.}~\bibnamefont {Caironi}}, \bibinfo {author} {\bibfnamefont {S.}~\bibnamefont {Trillo}}, \bibinfo {author} {\bibfnamefont {A.}~\bibnamefont {Piskarskas}},\ and\ \bibinfo {author} {\bibfnamefont {P.}~\bibnamefont {Di~Trapani}},\ }\bibfield  {title} {\bibinfo {title} {{Spatial versus temporal deterministic wave breakup of nonlinearly coupled light waves}},\ }\href@noop {} {\bibfield  {journal} {\bibinfo  {journal} {Phys. Rev. Lett.}\ }\textbf {\bibinfo {volume} {91}},\ \bibinfo {pages} {143905} (\bibinfo {year}
  {2003})}\BibitemShut {NoStop}%
\bibitem [{\citenamefont {Salerno}\ \emph {et~al.}(2004)\citenamefont {Salerno}, \citenamefont {Jedrkiewicz}, \citenamefont {Trull}, \citenamefont {Valiulis}, \citenamefont {Picozzi},\ and\ \citenamefont {Di~Trapani}}]{Salerno04}%
  \BibitemOpen
  \bibfield  {author} {\bibinfo {author} {\bibfnamefont {D.}~\bibnamefont {Salerno}}, \bibinfo {author} {\bibfnamefont {O.}~\bibnamefont {Jedrkiewicz}}, \bibinfo {author} {\bibfnamefont {J.}~\bibnamefont {Trull}}, \bibinfo {author} {\bibfnamefont {G.}~\bibnamefont {Valiulis}}, \bibinfo {author} {\bibfnamefont {A.}~\bibnamefont {Picozzi}},\ and\ \bibinfo {author} {\bibfnamefont {P.}~\bibnamefont {Di~Trapani}},\ }\bibfield  {title} {\bibinfo {title} {{Noise-seeded spatiotemporal modulation instability in normal dispersion}},\ }\href@noop {} {\bibfield  {journal} {\bibinfo  {journal} {Phys. Rev. E}\ }\textbf {\bibinfo {volume} {70}},\ \bibinfo {pages} {065603(R)} (\bibinfo {year} {2004})}\BibitemShut {NoStop}%
\bibitem [{\citenamefont {Schiek}\ and\ \citenamefont {Baronio}(2019)}]{Schiek19}%
  \BibitemOpen
  \bibfield  {author} {\bibinfo {author} {\bibfnamefont {R.}~\bibnamefont {Schiek}}\ and\ \bibinfo {author} {\bibfnamefont {F.}~\bibnamefont {Baronio}},\ }\bibfield  {title} {\bibinfo {title} {{Spatial Akhmediev breathers and modulation instability growth-decay cycles in a quadratic optical medium}},\ }\href@noop {} {\bibfield  {journal} {\bibinfo  {journal} {Phys. Rev. Research}\ }\textbf {\bibinfo {volume} {1}},\ \bibinfo {pages} {032036(R)} (\bibinfo {year} {2019})}\BibitemShut {NoStop}%
\bibitem [{\citenamefont {Schiek}(2021)}]{Schiek21}%
  \BibitemOpen
  \bibfield  {author} {\bibinfo {author} {\bibfnamefont {R.}~\bibnamefont {Schiek}},\ }\bibfield  {title} {\bibinfo {title} {{Excitation of nonlinear beams: from the linear Talbot effect through modulation instability to Akhmediev breathers}},\ }\href@noop {} {\bibfield  {journal} {\bibinfo  {journal} {Opt. Express}\ }\textbf {\bibinfo {volume} {29}},\ \bibinfo {pages} {15830} (\bibinfo {year} {2021})}\BibitemShut {NoStop}%
\bibitem [{\citenamefont {Deng}\ \emph {et~al.}(2022)\citenamefont {Deng}, \citenamefont {Zhang}, \citenamefont {Fan},\ and\ \citenamefont {Zhang}}]{Deng22}%
  \BibitemOpen
  \bibfield  {author} {\bibinfo {author} {\bibfnamefont {Z.}~\bibnamefont {Deng}}, \bibinfo {author} {\bibfnamefont {J.}~\bibnamefont {Zhang}}, \bibinfo {author} {\bibfnamefont {D.}~\bibnamefont {Fan}},\ and\ \bibinfo {author} {\bibfnamefont {L.}~\bibnamefont {Zhang}},\ }\bibfield  {title} {\bibinfo {title} {{Spatiotemporal doubly periodic waves in a phase-mismatched second-harmonic generation}},\ }\href@noop {} {\bibfield  {journal} {\bibinfo  {journal} {Opt. Lett.}\ }\textbf {\bibinfo {volume} {47}},\ \bibinfo {pages} {5557} (\bibinfo {year} {2022})}\BibitemShut {NoStop}%
\bibitem [{\citenamefont {Trillo}\ and\ \citenamefont {Baronio}(2023)}]{Trillo23}%
  \BibitemOpen
  \bibfield  {author} {\bibinfo {author} {\bibfnamefont {S.}~\bibnamefont {Trillo}}\ and\ \bibinfo {author} {\bibfnamefont {F.}~\bibnamefont {Baronio}},\ }\bibfield  {title} {\bibinfo {title} {{From regular to pseudo-stochastic recurrence of modulation instability in cascaded second-harmonic generation}},\ }\href@noop {} {\bibfield  {journal} {\bibinfo  {journal} {Opt. Lett.}\ }\textbf {\bibinfo {volume} {48}},\ \bibinfo {pages} {1284} (\bibinfo {year} {2023})}\BibitemShut {NoStop}%
\bibitem [{\citenamefont {Stegeman}(1997)}]{cascading}%
  \BibitemOpen
  \bibfield  {author} {\bibinfo {author} {\bibfnamefont {G.~I.}\ \bibnamefont {Stegeman}},\ }\bibfield  {title} {\bibinfo {title} {{$\chi^{(2)}$ cascading: nonlinear phase shifts}},\ }\href@noop {} {\bibfield  {journal} {\bibinfo  {journal} {Quantum Semiclass. Opt.}\ }\textbf {\bibinfo {volume} {9}},\ \bibinfo {pages} {139} (\bibinfo {year} {1997})}\BibitemShut {NoStop}%
\bibitem [{\citenamefont {Akhmediev}\ \emph {et~al.}(1987)\citenamefont {Akhmediev}, \citenamefont {Eleonskii},\ and\ \citenamefont {Kulagin}}]{Akh87}%
  \BibitemOpen
  \bibfield  {author} {\bibinfo {author} {\bibfnamefont {N.~N.}\ \bibnamefont {Akhmediev}}, \bibinfo {author} {\bibfnamefont {V.~M.}\ \bibnamefont {Eleonskii}},\ and\ \bibinfo {author} {\bibfnamefont {N.~E.}\ \bibnamefont {Kulagin}},\ }\bibfield  {title} {\bibinfo {title} {{Exact first-order solutions of the nonlinear Schr\"odinger equation}},\ }\href@noop {} {\bibfield  {journal} {\bibinfo  {journal} {Theor. Math. Phys. (USSR)}\ }\textbf {\bibinfo {volume} {72}},\ \bibinfo {pages} {809} (\bibinfo {year} {1987})}\BibitemShut {NoStop}%
\bibitem [{\citenamefont {Grinevich}\ and\ \citenamefont {Santini}(2018)}]{GS18}%
  \BibitemOpen
  \bibfield  {author} {\bibinfo {author} {\bibfnamefont {P.~G.}\ \bibnamefont {Grinevich}}\ and\ \bibinfo {author} {\bibfnamefont {P.~M.}\ \bibnamefont {Santini}},\ }\bibfield  {title} {\bibinfo {title} {{The exact rogue wave recurrence in the NLS periodic setting via matched asymptotic expansions, for 1 and 2 unstable modes}},\ }\href@noop {} {\bibfield  {journal} {\bibinfo  {journal} {Phys. Lett. A}\ }\textbf {\bibinfo {volume} {382}},\ \bibinfo {pages} {973} (\bibinfo {year} {2018})}\BibitemShut {NoStop}%
\bibitem [{\citenamefont {Conforti}\ \emph {et~al.}(2020)\citenamefont {Conforti}, \citenamefont {Mussot}, \citenamefont {Kudlinski}, \citenamefont {Trillo},\ and\ \citenamefont {Akhmediev}}]{Conforti20}%
  \BibitemOpen
  \bibfield  {author} {\bibinfo {author} {\bibfnamefont {M.}~\bibnamefont {Conforti}}, \bibinfo {author} {\bibfnamefont {A.}~\bibnamefont {Mussot}}, \bibinfo {author} {\bibfnamefont {A.}~\bibnamefont {Kudlinski}}, \bibinfo {author} {\bibfnamefont {S.}~\bibnamefont {Trillo}},\ and\ \bibinfo {author} {\bibfnamefont {N.}~\bibnamefont {Akhmediev}},\ }\bibfield  {title} {\bibinfo {title} {{Doubly periodic solutions of the focusing nonlinear Schr\"odinger equation: recurrence, period doubling and amplication outside the conventional modulation instability band}},\ }\href@noop {} {\bibfield  {journal} {\bibinfo  {journal} {Phys. Rev. A}\ }\textbf {\bibinfo {volume} {101}},\ \bibinfo {pages} {023843} (\bibinfo {year} {2020})}\BibitemShut {NoStop}%
\bibitem [{\citenamefont {Baronio}(2017)}]{Baronio17}%
  \BibitemOpen
  \bibfield  {author} {\bibinfo {author} {\bibfnamefont {F.}~\bibnamefont {Baronio}},\ }\bibfield  {title} {\bibinfo {title} {{Akhmediev Breathers and Peregrine Solitary Waves in a Quadratic Medium}},\ }\href@noop {} {\bibfield  {journal} {\bibinfo  {journal} {Opt. Lett.}\ }\textbf {\bibinfo {volume} {42}},\ \bibinfo {pages} {1756} (\bibinfo {year} {2017})}\BibitemShut {NoStop}%
\bibitem [{\citenamefont {Di~Trapani}\ \emph {et~al.}(1998)\citenamefont {Di~Trapani}, \citenamefont {Caironi}, \citenamefont {Valiulis}, \citenamefont {Dubietis}, \citenamefont {Danielius},\ and\ \citenamefont {Piskarskas}}]{DiTrapani98}%
  \BibitemOpen
  \bibfield  {author} {\bibinfo {author} {\bibfnamefont {P.}~\bibnamefont {Di~Trapani}}, \bibinfo {author} {\bibfnamefont {D.}~\bibnamefont {Caironi}}, \bibinfo {author} {\bibfnamefont {G.}~\bibnamefont {Valiulis}}, \bibinfo {author} {\bibfnamefont {A.}~\bibnamefont {Dubietis}}, \bibinfo {author} {\bibfnamefont {R.}~\bibnamefont {Danielius}},\ and\ \bibinfo {author} {\bibfnamefont {A.}~\bibnamefont {Piskarskas}},\ }\bibfield  {title} {\bibinfo {title} {{Observation of Temporal Solitons in Second-Harmonic Generation with Tilted Pulses}},\ }\href@noop {} {\bibfield  {journal} {\bibinfo  {journal} {Phys. Rev. Lett.}\ }\textbf {\bibinfo {volume} {81}},\ \bibinfo {pages} {570} (\bibinfo {year} {1998})}\BibitemShut {NoStop}%
\bibitem [{\citenamefont {Jankowski}\ \emph {et~al.}(2020)\citenamefont {Jankowski}, \citenamefont {Langrock}, \citenamefont {Desiatov}, \citenamefont {Marandi}, \citenamefont {Wang}, \citenamefont {Zhang}, \citenamefont {Phillips}, \citenamefont {Lon\v{c}ar},\ and\ \citenamefont {Fejer}}]{Jankowski20}%
  \BibitemOpen
  \bibfield  {author} {\bibinfo {author} {\bibfnamefont {M.}~\bibnamefont {Jankowski}}, \bibinfo {author} {\bibfnamefont {C.}~\bibnamefont {Langrock}}, \bibinfo {author} {\bibfnamefont {B.}~\bibnamefont {Desiatov}}, \bibinfo {author} {\bibfnamefont {A.}~\bibnamefont {Marandi}}, \bibinfo {author} {\bibfnamefont {C.}~\bibnamefont {Wang}}, \bibinfo {author} {\bibfnamefont {M.}~\bibnamefont {Zhang}}, \bibinfo {author} {\bibfnamefont {C.~R.}\ \bibnamefont {Phillips}}, \bibinfo {author} {\bibfnamefont {M.}~\bibnamefont {Lon\v{c}ar}},\ and\ \bibinfo {author} {\bibfnamefont {M.~M.}\ \bibnamefont {Fejer}},\ }\bibfield  {title} {\bibinfo {title} {Ultrabroadband nonlinear optics in nanophotonic periodically poled lithium niobate waveguides},\ }\href {https://doi.org/10.1364/OPTICA.7.000040} {\bibfield  {journal} {\bibinfo  {journal} {Optica}\ }\textbf {\bibinfo {volume} {7}},\ \bibinfo {pages} {40} (\bibinfo {year} {2020})}\BibitemShut {NoStop}%
\bibitem [{\citenamefont {Kumar}\ \emph {et~al.}(2022)\citenamefont {Kumar}, \citenamefont {Younesi}, \citenamefont {Saravi}, \citenamefont {Setzpfandt},\ and\ \citenamefont {Pertsch}}]{Kumar22}%
  \BibitemOpen
  \bibfield  {author} {\bibinfo {author} {\bibfnamefont {P.}~\bibnamefont {Kumar}}, \bibinfo {author} {\bibfnamefont {M.}~\bibnamefont {Younesi}}, \bibinfo {author} {\bibfnamefont {S.}~\bibnamefont {Saravi}}, \bibinfo {author} {\bibfnamefont {F.}~\bibnamefont {Setzpfandt}},\ and\ \bibinfo {author} {\bibfnamefont {T.}~\bibnamefont {Pertsch}},\ }\bibfield  {title} {\bibinfo {title} {{Group-index-matched frequency conversion in lithium niobate on insulator waveguides}},\ }\href@noop {} {\bibfield  {journal} {\bibinfo  {journal} {Front. Photonics}\ }\textbf {\bibinfo {volume} {3}},\ \bibinfo {pages} {951949} (\bibinfo {year} {2022})}\BibitemShut {NoStop}%
\bibitem [{\citenamefont {Trillo}\ \emph {et~al.}(1992)\citenamefont {Trillo}, \citenamefont {Wabnitz}, \citenamefont {Chisari},\ and\ \citenamefont {Cappellini}}]{Trillo92a}%
  \BibitemOpen
  \bibfield  {author} {\bibinfo {author} {\bibfnamefont {S.}~\bibnamefont {Trillo}}, \bibinfo {author} {\bibfnamefont {S.}~\bibnamefont {Wabnitz}}, \bibinfo {author} {\bibfnamefont {R.}~\bibnamefont {Chisari}},\ and\ \bibinfo {author} {\bibfnamefont {G.}~\bibnamefont {Cappellini}},\ }\bibfield  {title} {\bibinfo {title} {{Two wave mixing in a quadratic nonlinear medium: bifurcations, spatial instabilities, and chaos}},\ }\href@noop {} {\bibfield  {journal} {\bibinfo  {journal} {Opt. Lett.}\ }\textbf {\bibinfo {volume} {17}},\ \bibinfo {pages} {637} (\bibinfo {year} {1992})}\BibitemShut {NoStop}%
\bibitem [{\citenamefont {Byrd}\ and\ \citenamefont {Friedman}(1971)}]{elliptic}%
  \BibitemOpen
  \bibfield  {author} {\bibinfo {author} {\bibfnamefont {P.~F.}\ \bibnamefont {Byrd}}\ and\ \bibinfo {author} {\bibfnamefont {M.~D.}\ \bibnamefont {Friedman}},\ }\href@noop {} {\emph {\bibinfo {title} {{Handbook of Elliptic Integrals for Engineers and Scientists}}}}\ (\bibinfo  {publisher} {Springer-Verlag},\ \bibinfo {address} {Berlin},\ \bibinfo {year} {1971})\BibitemShut {NoStop}%
\bibitem [{\citenamefont {Armstrong}\ \emph {et~al.}(1962)\citenamefont {Armstrong}, \citenamefont {Bloembergen}, \citenamefont {Ducuing},\ and\ \citenamefont {Pershan}}]{Armstrong62}%
  \BibitemOpen
  \bibfield  {author} {\bibinfo {author} {\bibfnamefont {J.~A.}\ \bibnamefont {Armstrong}}, \bibinfo {author} {\bibfnamefont {N.}~\bibnamefont {Bloembergen}}, \bibinfo {author} {\bibfnamefont {J.}~\bibnamefont {Ducuing}},\ and\ \bibinfo {author} {\bibfnamefont {P.~S.}\ \bibnamefont {Pershan}},\ }\bibfield  {title} {\bibinfo {title} {{Interactions between light waves in a nonlinear dielectric}},\ }\href@noop {} {\bibfield  {journal} {\bibinfo  {journal} {Phys. Rev.}\ }\textbf {\bibinfo {volume} {127}},\ \bibinfo {pages} {1918} (\bibinfo {year} {1962})}\BibitemShut {NoStop}%
\bibitem [{\citenamefont {Mersman}(1970)}]{Mersman1970}%
  \BibitemOpen
  \bibfield  {author} {\bibinfo {author} {\bibfnamefont {W.~A.}\ \bibnamefont {Mersman}},\ }\bibfield  {title} {\bibinfo {title} {{A new algorithm for the Lie transformation}},\ }\href@noop {} {\bibfield  {journal} {\bibinfo  {journal} {Celestial Mechanics}\ }\textbf {\bibinfo {volume} {3}},\ \bibinfo {pages} {81} (\bibinfo {year} {1970})}\BibitemShut {NoStop}%
\bibitem [{\citenamefont {Cary}(1981)}]{CaryReview1981}%
  \BibitemOpen
  \bibfield  {author} {\bibinfo {author} {\bibfnamefont {J.~R.}\ \bibnamefont {Cary}},\ }\bibfield  {title} {\bibinfo {title} {{Lie transform perturbation theory for Hamiltonian systems}},\ }\href@noop {} {\bibfield  {journal} {\bibinfo  {journal} {Physics Reports}\ }\textbf {\bibinfo {volume} {79}},\ \bibinfo {pages} {129} (\bibinfo {year} {1981})}\BibitemShut {NoStop}%
\bibitem [{\citenamefont {Sanders}\ \emph {et~al.}(2007)\citenamefont {Sanders}, \citenamefont {Verhulst},\ and\ \citenamefont {Murdock}}]{VerhulstBook}%
  \BibitemOpen
  \bibfield  {author} {\bibinfo {author} {\bibfnamefont {J.~A.}\ \bibnamefont {Sanders}}, \bibinfo {author} {\bibfnamefont {F.}~\bibnamefont {Verhulst}},\ and\ \bibinfo {author} {\bibfnamefont {J.}~\bibnamefont {Murdock}},\ }\href@noop {} {\emph {\bibinfo {title} {{Averaging Methods in Nonlinear Dynamical Systems}}}},\ \bibinfo {edition} {2nd}\ ed.\ (\bibinfo  {publisher} {Springer-Verlag},\ \bibinfo {address} {Berlin},\ \bibinfo {year} {2007})\BibitemShut {NoStop}%
\end{thebibliography}
\end{document}